%% file: main.tex
\documentclass[sigplan,10pt,screen]{acmart}
\renewcommand\footnotetextcopyrightpermission[1]{}
\usepackage{tikz}
\usepackage{amsmath}
\usepackage{pifont}
\usepackage{comment}

\usepackage[normalem]{ulem}

\usepackage{xspace}
\usepackage{algorithm}
\usepackage[noend]{algpseudocode}
\usepackage{wrapfig}
\usepackage{subfig}
\usepackage{multirow}
\usepackage{multicol}
\usepackage{enumitem}

\newcommand{\proj}{Kudu\xspace}

\newcommand{\red}[1]{\textcolor{black}{#1}}
\newcommand{\blue}[1]{\textcolor{black}{#1}}
\newcommand{\rev}[1]{\textcolor{black}{#1}}
\newcommand{\revision}[1]{\textcolor{black}{#1}}

\newcommand{\maxspeedupkautominegthinker}[0]{$75.5\times$\xspace}
\newcommand{\maxspeedupkgraphpigthinker}[0]{$75.5\times$\xspace}
\newcommand{\avgspeedupkautominegthinker}[0]{$17.7\times$\xspace}
\newcommand{\avgspeedupkgraphpigthinker}[0]{$20.3\times$\xspace}
\newcommand{\maxspeedupgthinker}[0]{$75.5\times$\xspace}
\newcommand{\avgspeedupgthinker}[0]{$19.0\times$\xspace}

\begin{document}

\title{Kudu: An Efficient and Scalable Distributed Graph Pattern Mining Engine}

\author{Jingji Chen}
\email{chen3385@purdue.edu}
\affiliation{%
  \institution{Purdue University}
  \city{West Lafayette, IN}
  \country{USA}}

\author{Xuehai Qian}
\email{qian214@purdue.edu}
\affiliation{%
  \institution{Purdue University}
  \city{West Lafayette, IN}
  \country{USA}}

\begin{abstract}
This paper proposes \proj, a distributed
execution engine with a well-defined abstraction
that can be integrated
with existing single-machine graph pattern mining (GPM) systems
\rev{to provide efficiency and scalability at the same time}.
The key novelty is the extendable embedding abstraction
which can express pattern 
enumeration algorithms, allow 
fine-grained task scheduling, and enable low-cost
GPM-specific data reuse \red{to reduce communication cost}.
The effective BFS-DFS hybrid exploration generates 
sufficient concurrent tasks for communication-computation overlapping
with bounded memory consumption.
Two scalable distributed GPM systems 
are implemented by porting Automine and GraphPi on \proj.
Our evaluation shows that \proj based systems significantly outperform state-of-the-art distributed GPM systems with partitioned graphs
\red{by up to \maxspeedupgthinker (on average \avgspeedupgthinker)}, 
achieve \red{similar or even better} performance compared with the fastest distributed GPM systems 
with replicated graph, 
and scale to \rev{massive graphs with more than one hundred billion edges with a commodity cluster.}
\end{abstract}

\settopmatter{printfolios=true}
\settopmatter{printacmref=false}
\maketitle
\pagestyle{plain}

\input{intro.tex}

\input{back.tex}

\input{problem.tex}

\input{extendable.tex}

\input{hybrid.tex}

\input{reduction.tex}

\input{impl.tex}

\input{eval.tex}
\input{conc.tex}


\bibliographystyle{ACM-Reference-Format}
\bibliography{graph_mining}

\end{document}

%% file: intro.tex
\section{Introduction}
\label{sec:intro}

Graph pattern mining (GPM)~\cite{teixeira2015arabesque},
an important graph processing
workload, is widely used in various 
applications~\cite{ma2009insights,schmidt2011efficient,wu2018software,valverde2005network,staniford1996grids,juszczyszyn2011motif,becchetti2008efficient}.
Given an input graph, GPM enumerates all its subgraphs isomorphic to some user-defined pattern(s), known as {\em embeddings},
and processes them to extract useful information.
GPM application is computation intensive
due to the need to enumerate an extremely large
number of subgraphs. 
For example, there are more than 30 trillion edge-induced 6-chain embeddings on WikiVote~\cite{leskovec2010signed}---a tiny graph with only 7K vertices. 
On the other side, the size of the graphs in real-world applications
is continuously increasing. 
The importance and complexity of GPM
applications give rise to the
recent general GPM systems~\cite{teixeira2015arabesque,wang2018rstream,dias2019fractal,mawhirter2019automine,jamshidi2020peregrine,chen2020pangolin,shi2020graphpi,bindschaedler2021tesseract,chen2018g,yan2020g,chen2021sandslash}.
\rev{In order to process large graphs efficiently,}
a GPM system 
should {\em scale with both the computation and
memory resources with good execution efficiency}.

GPM can be performed with two approaches:
(1) {\em pattern-oblivious} method---enumerating all subgraphs up to the pattern size and performing expensive isomorphic checks---used in early GPM systems such as Arabesque~\cite{teixeira2015arabesque}, 
Fractal~\cite{dias2019fractal}, and 
RStream~\cite{wang2018rstream}; and 
(2) {\em pattern-aware} enumeration method---generating the embeddings
that match the patterns by construction without
isomorphism checks---used in recent GPM systems 
such as Automine~\cite{jamshidi2020peregrine}, GraphZero~\cite{mawhirter2021graphzero},
Peregrine~\cite{jamshidi2020peregrine}, and
GraphPi~\cite{shi2020graphpi}. 
This paper focuses on the pattern-aware enumeration due to 
its significantly better performance.

\begin{figure}[t]
    \centering
    \includegraphics[width=\linewidth]{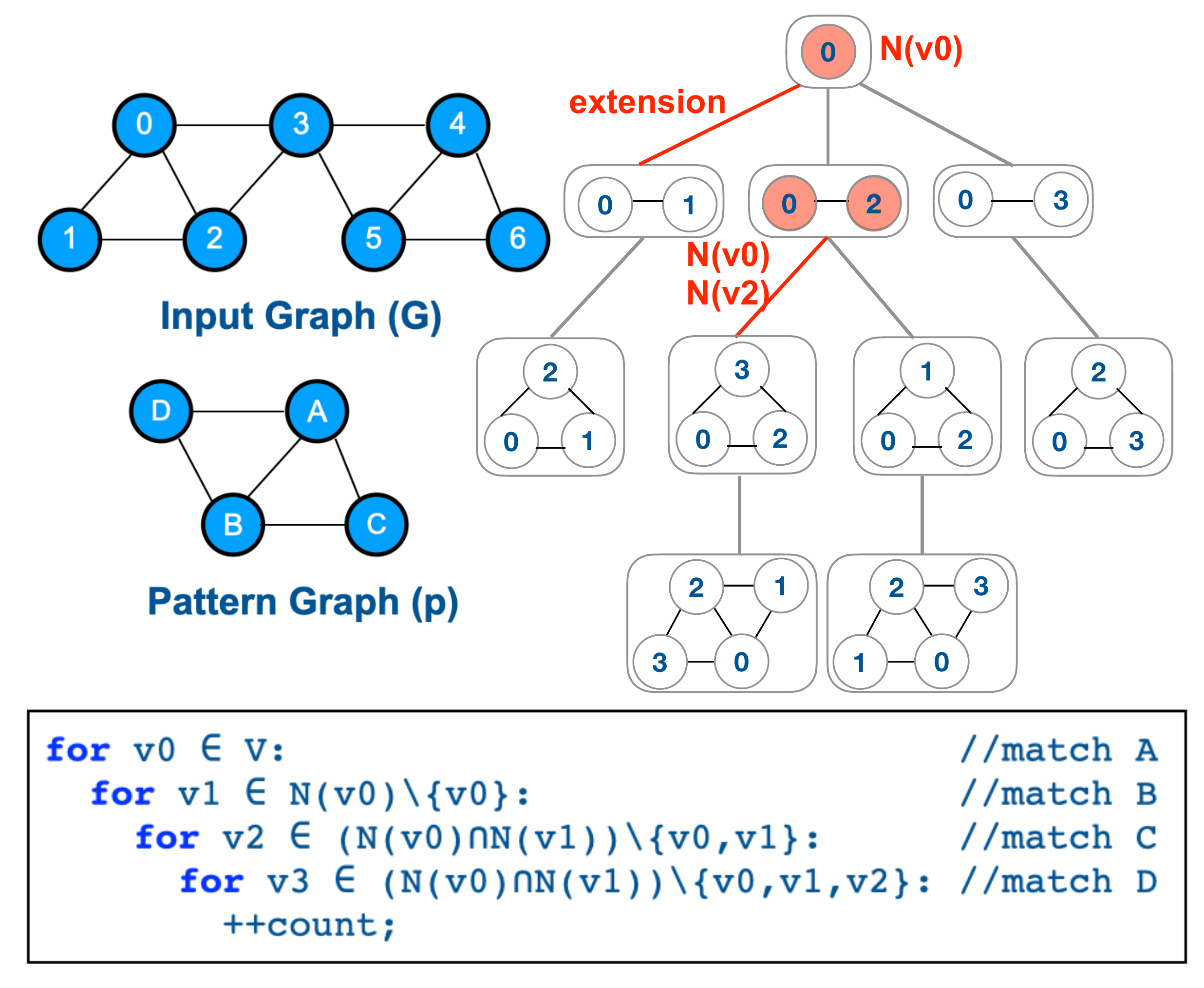}
    \caption{Pattern-aware Method and Embedding Tree}
    \label{fig:enum_tree}
\end{figure}

Figure~\ref{fig:enum_tree} shows an example of the pattern-aware
enumeration method. The nested loops construct the pattern embeddings
in matching order (A,B,C,D), each loop level matches a vertex
in the pattern graph. 
Such nested loops are executed from each vertex in the input graph. 
For each vertex $v0$, the code explores a tree structure with 
$v0$ as the root. Each tree node corresponds to a partially-constructed
embedding or an embedding matching the pattern.
Each tree level corresponds to a loop level.
For example, the third loop (matching C)
generates two tree nodes corresponding to 
three 2-vertex partially-constructed embedding
($v0$,$v2$,$v3$) and ($v0$,$v2$,$v1$) from ($v0$,$v2$).
In general, a node needs to access an edge list ($N(v2)$ in above example) and 
perform intersections to generate a children.
$N(v)$ is the edge list (neighbors) of $v$.
We can see that the shape of the tree depends on the input graph. 
The tree leaves can be either embeddings matching the pattern, or 
subgraphs that cannot be extended further to match the pattern. 

In a nutshell, the computation of GPM is to explore 
a massive number of trees (equal to the number of 
vertices in the input graph) based on the pattern graph and the input graph.
We call such trees as {\em embedding trees}.
The current GPM systems can be classified into two types based 
on whether remote data is accessed during the exploration of each
embedding tree.

The first does not access remote edge lists during enumeration,
including {\em single-machine GPM systems}, such as
AutoMine~\cite{mawhirter2019automine}, Peregrine~\cite{jamshidi2020peregrine}, GraphZero~\cite{mawhirter2021graphzero}, Pangolin~\cite{chen2020pangolin}, and 
Sandslash~\cite{chen2021sandslash},
and {\em distributed 
GPM systems with replicated graphs}, such as 
Fractal~\cite{dias2019fractal},
Arabesque~\cite{teixeira2015arabesque}, 
and GraphPi~\cite{shi2020graphpi}.
The single-machine systems, which assume
the input graph can fit into the memory, can support
rudimentary out-of-core mode with
memory mapped I/O to accommodate large graphs
stored in disk but it may lead to
poor performance for complicated patterns.
RStream~\cite{wang2018rstream} is a
specific out-of-core system that optimizes disk accesses, 
but it can only use the computation resource of one machine. 
The distributed systems scale with computation resources but 
the graph size is restricted by the memory of each machine.

The second type 
is {\em distributed GPM systems with partitioned graphs},
including G-miner~\cite{chen2018g} and its successor G-thinker~\cite{yan2020g}, which access remote edge lists
during enumeration.
G-thinker and G-miner are designed with the 
``moving data to computation'' policy.
A recent distributed graph-querying system aDFS~\cite{trigonakis2021adfs}
performs GPM tasks using an expressive SQL-like  query language. 
It follows the ``moving computation to data'' policy 
that prevents many communication optimizations, leading to low performance for GPM tasks. 
For example, aDFS takes roughly 1000 seconds for triangle \red{(size-3 complete subgraph)} counting on the Friendster graph with 224 cores~\cite{trigonakis2021adfs} while  AutoMine~\cite{mawhirter2019automine} only takes roughly 400 seconds with 16 cores.

Tesseract~\cite{bindschaedler2021tesseract} is a distributed GPM system optimized for dynamic graph updates, which stores the graph with
a disaggregated key-value store. We do not consider it as a
system with partitioned graphs. Its performance is not 
competitive compared to the state-of-the-art GPM systems:
Tesseract takes 1.9 hours to mine \rev{all size-4} patterns on the LiveJournal~\cite{leskovec2009community} graph with eight 16-core machines~\cite{bindschaedler2021tesseract} while
GraphPi can perform the same task in 279 seconds with one machine.

\begin{figure}[t]
    \centering
    \includegraphics[width=\linewidth]{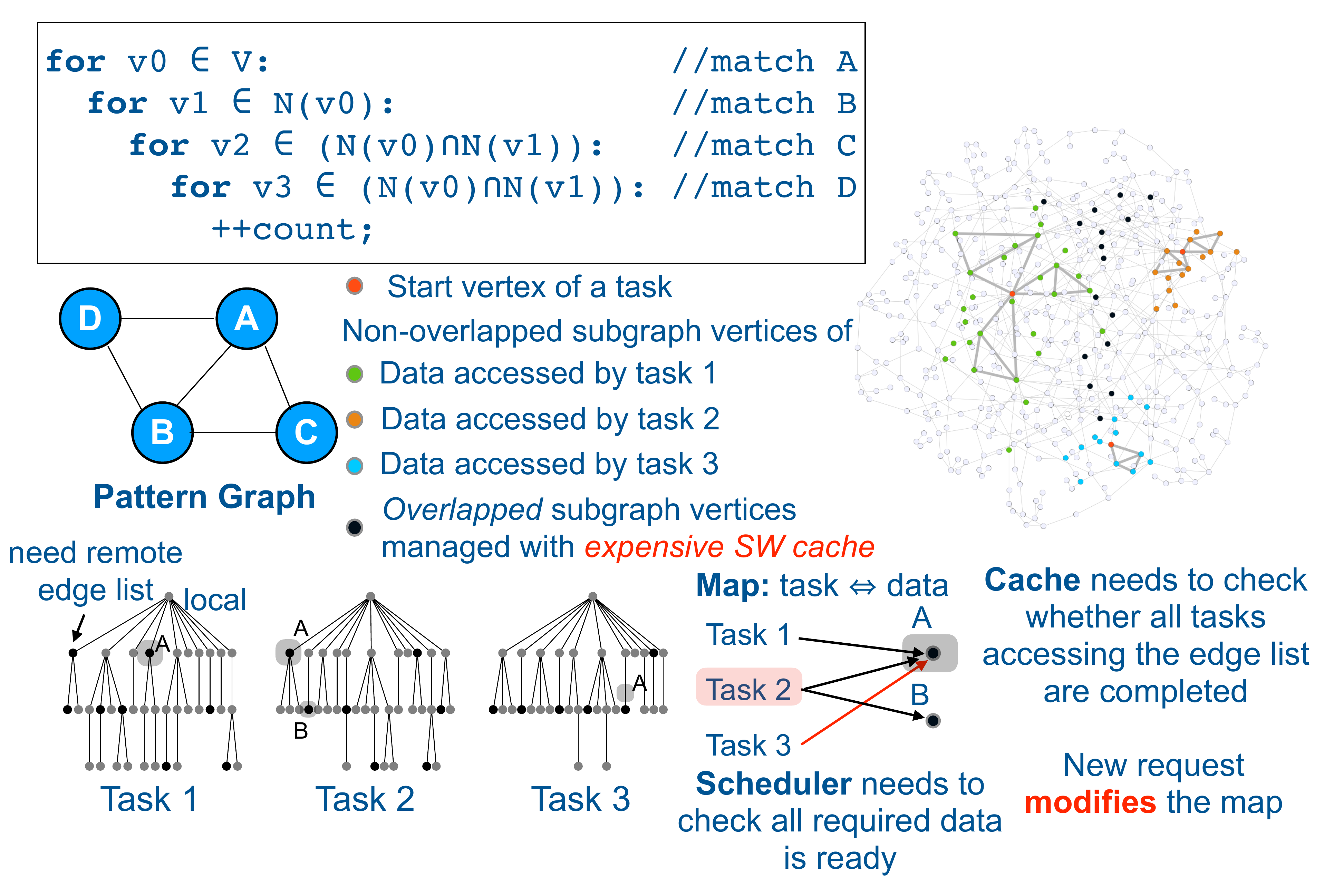}
    \caption{G-thinker: High Scheduler and Cache Overhead }
    \label{fig:gt_problem}
\end{figure}

{\bf Problem of G-thinker:} 
We run experiments using the publicly available implementation.
It \red{takes 27.2s seconds for triangle counting on the Patents~\cite{leskovec2005graphs} graph on an 8-node cluster (8 cores per node) while a single-thread version finishes within 6.2 seconds~\cite{mawhirter2019automine}}. 
The low performance of G-thinker is due to two reasons
shown in Figure~\ref{fig:gt_problem}.

\begin{itemize}[leftmargin=*]
\item {\bf Limited parallelism.}
    \red{Each task of G-thinker explores an entire embedding tree.}
    Each task of G-thinker explores an entire embedding tree.
    Before the computation, the task fetches a k-hop subgraph \red{containing all necessary data for the tree exploration}.
    Because the size of the k-hop subgraph increases quickly with k,
    the memory consumption of each task is large.
    Thus, the number of embedding trees that can be 
concurrently explored is reduced, e.g., \red{about 150-300} trees
for \red{triangle counting} on the \red{Patents graph}.
\item {\bf General but expensive software cache and scheduler.}  
To enable data reuse \red{among tasks} and avoid 
fetching the same edge list multiple 
times, 
G-thinker implements a software cache for remote edge lists management, 
which maintain a {\em map between tasks and 
\red{the dependent} edge lists in cache, incurring high computation overhead}. 
When a task requests an edge list of
a vertex, the map needs to be {\em updated}.
The scheduler periodically checks whether the edge lists needed
by each task is ready. 
The cache periodically checks whether the tasks accessing an edge list
are all completed, if so, it can be garbage collected. 

\end{itemize}

\begin{figure}[t]
    \centering
    \includegraphics[width=\linewidth]{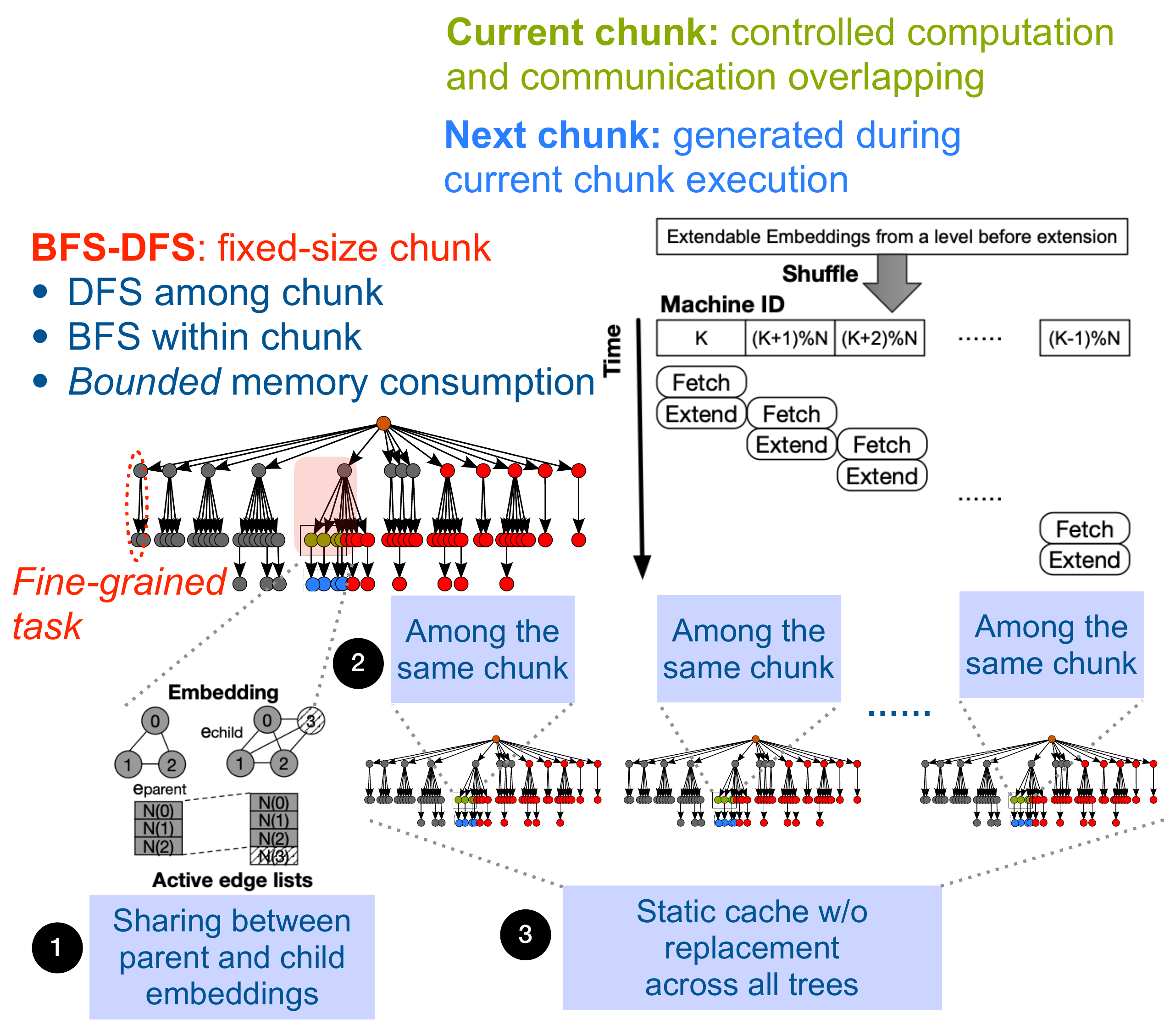}
    \caption{Benefits of Extendable Embedding Abstraction}
    \label{fig:benefits}
\end{figure}

{\bf Key idea: extendable embedding.} We propose 
a novel abstraction that enables the efficient distributed execution of
GPM algorithm. 
The key insight is to consider {\em each extension as a fine-grained 
task}. 
In an embedding tree (refer to Figure~\ref{fig:enum_tree}), 
each task corresponds 
to {\em an edge in the tree from parent to its child}, 
e.g., an edge between 
$(v0,v2)$ (parent) and $(v0,v2,v1)$ (child).
{\em Extendable embedding} is the abstraction to realize such 
fine-grained tasks: an extendable embedding can be considered as
a tree node---a partially-constructed embedding---together with 
some active edge lists, which are needed to perform the extension, i.e., generating the child. 
The extendable embedding for the tree node $(v0,v2)$ is 
$(v0,v2)$ plus $N(v0)$ and $N(v1)$, because the new vertex 
($v1$ or $v3$) extended is identified by the 
intersection of $N(v0)$ and $N(v1)$.
Extendable embedding is a well-defined abstraction that 
specifies the {\em computation} and the dependent {\em data}
based on which the computation is performed.
{\em A fine-grained task performs the extension based on an extendable 
embedding after all active edge lists are fetched to local machine}.
In contrast, each task in G-thinker performs the exploration of a
whole tree.
The abstraction provides two important benefits illustrated in Figure~\ref{fig:benefits}.

{\bf Benefit 1: enabling flexible BFS-DFS hybrid exploration.}
The fine-grained task enabled by extendable embedding allows
the flexible scheduling to take the best of BFS (high parallelism)
and DFS (low memory consumption). 
We propose a BFS-DFS hybrid exploration with {\em fixed-size extendable
embedding chunks
that performs DFS among chunks and BFS within a chunk}.
During the execution, the {\em current} chunk fills the {\em next} chunk 
until it is full. 
The hybrid approach bounds memory consumption because
{\em for each tree, we only need to allocate the memory of k chunks, 
where k is the level of tree nodes being explored}. 
Before processing a chunk, we can 
shuffle the active edge lists based on the machine ID, and perform
the computation and communication in a \red{lightweight} pipelined manner to 
maximize the chance of overlapping. 
While BFS-DFS exploration is also used in aDFS~\cite{trigonakis2021adfs}, 
that system uses ``moving computation to data'' policy
with poor performance for GPM. 

{\bf Benefit 2: GPM-specific low-cost data reuse.}
Compared to the general software cache used in G-thinker, the 
fine-grained tasks enable three types of effective data reuse leveraging GPM algorithm properties. 
First, the data reuse between parent and child (\ding{182}):
the child is extended with a new vertex from the parent, reuse of the
shared data can be realized by pointers in the child to the parent.
Second, data sharing among extendable embeddings in the same chunk (\ding{183}) can be realized by a \red{simplified} hash table.
Third, the hot graph data can be kept in local machine by a static cache
(\ding{184}).
Unlike G-thinker, 
the static cache only inserts fetched edge list
by certain criteria until it is full but {\em never evicts data}. 
It only reduces communication and remote edge lists access latency,
but does not incur high overhead because 
\red{the expensive bookkeeping that maintains the mapping between tasks and dependent edge lists}
is not needed.

{\bf The system: } We built {\em \proj}, \rev{the first} distributed
execution engine with a well-defined abstraction 
that can be conveniently integrated
with existing single-machine GPM systems.
This approach {\em keeps the user-facing 
programming interface
of existing systems unmodified}, which typically 
specifies the pattern to mine,
and only requires modest changes to the 
existing GPM systems' implementations.
Specifically, a GPM system developer can implement
the pattern-specific subgraph enumeration, i.e., 
a sequence of extensions in the embedding tree shown 
in Figure~\ref{fig:enum_tree}, in 
a single \texttt{EXTEND} function, which is the sole interface
between a single-machine GPM system and \proj.
Internally, \proj implements a scheduler and execution model 
to realize the hybrid scheduling.

{\bf Experimental result highlights: }
We built two scalable \rev{and efficient} distributed GPM systems, k-Automine and k-GraphPi, by porting Automine~\cite{mawhirter2019automine} and GraphPi~\cite{shi2020graphpi} on top of \proj,
with porting cost of roughly 500 lines of code per system.
k-Automine and k-GraphPi significantly outperform G-thinker by up to \maxspeedupkautominegthinker  and \maxspeedupkgraphpigthinker  (on average \avgspeedupkautominegthinker and \avgspeedupkgraphpigthinker), respectively.
Compared to GraphPi, the fastest distributed GPM system with replicated graph,
\proj based systems show similar or even better performance compared 
and scale to massive graphs \emph{with more than one hundred billion edges} that replication based systems cannot handle.
More impressively, k-Automine even outperforms a state-of-the-art CPU-based distributed triangle counting implementation~\cite{pearce2017triangle} targeting large graphs. 

{\bf Key lesson:} Unless shared through the static cache,
there is no sharing between two extendable
embeddings of different trees, or
different levels of the same tree
if one is not the decedent of another.
In contrast, G-thinker allows the sharing of all 
edge lists in the cache, which indeed translates
to {\em less communication} between machines than \proj. 
However, our approach leverages GPM-specific properties 
to enable low-cost data sharing and 
avoids the high overhead of G-thinker.
\emph{Our results show that on average \proj incurs about 3$\times$
communication of G-thinker but achieves average
speedups of \avgspeedupgthinker (up to \maxspeedupgthinker)}. 
{\em Not solely focusing on one aspect and effectively exploring the 
trade-offs between overhead and communication amount, is the most 
important takeaway}.



%% file: back.tex
\section{Background}
\label{sec:back}

\subsection{Graph Mining Problem}
\label{sec:mining_basic}


A graph $G$ is defined as $G=(V,E)$, in which $V$ and $E$ are the set of graph vertices and edges.
Similar to other systems~\cite{teixeira2015arabesque,chen2020pangolin,mawhirter2019automine,jamshidi2020peregrine}, this paper considers undirected graphs, 
while the techniques are also 
applicable to directed graphs.
We denote the vertex set containing $v$'s neighbors as $N(v)$.
Each vertex or edge may have a label, and can be represented by a mapping $f_L:V\cup E\rightarrow L$ where $L$ is the label set.
Currently, \proj supports vertex labels, but
the edge label support can be added without
fundamental difficulty. 
Two graphs $G_0=(V_0, E_0)$ and $G_1=(V_1, E_1)$ are isomorphic iff. there exists a bijective function $f:V_0\rightarrow V_1$ such that $(u,v)\in E_0 \iff (f(u), f(v))\in E_1$.
Intuitively, isomorphic graphs \rev{have} the same structure.

Given an input graph $G$, 
a GPM task discovers and processes $G$'s subgraphs that are isomorphic with a user-specified \textit{pattern} $p$, which
is a small connected graph reflecting some application-specific knowledge. 
The subgraphs matching (isomorphic with) $p$ is called $p$'s {\em embeddings}.
In Figure~\ref{fig:enum_tree}, $G$'s subgraphs $e_1$ and $e_2$ are isomorphic 
to the pattern graph $p$.

%% file: problem.tex
\subsection{Distributed Graph Pattern Mining}
\label{sec:problem}

Graph partitioning divides
graph data into multiple partitions, 
each of which is stored in the memory 
of one machine of a distributed cluster.
Thus, the application can scale with both memory
and compute resources. 
In this paper, we use 
1-D graph partitioning: the vertices set $V$ of the input graph is partitioned into $N$ parts $V_0, V_1, \ldots, V_{N-1}$, where $N$ is the number of machines.
Machine $i$ ($0\le i<N$) maintains all graph data related to $V_i$---all edges with at least one endpoint in $V_i$---in its memory.  
To ensure balanced data distribution, similar to previous systems~\cite{malewicz2010pregel,yan2020g}, the graph partitioning is determined by a hash function $H(v)$ that maps a vertex $v$ to its partition ID---an integer between $0$ and $N-1$.
Figure~\ref{fig:partition_gpm}
shows an example of input graph partitioning among three machines.
Each machine contains 
a subset of vertices
mapped to it and all edges
that are connected to them.

With graph partition, the graph data needed in 
enumeration may not exist in the 
memory of the local machine---incurring remote data accesses
and communication overhead. 
Such overhead can be 
exaggerated in the skewed graphs---the real-world graphs that follow the ``power law''.
Specifically, \rev{since} the high-degree vertices \red{with long edge lists} belong
to much more subgraphs \rev{and are more frequently accessed}, 
the fetch of their edge
lists can lead to a tremendous amount of communication.

\subsection{Two Strategies for Distributed GPM}
\label{sec:limit}


\begin{figure}
    \centering
    \includegraphics[width=.9\linewidth]{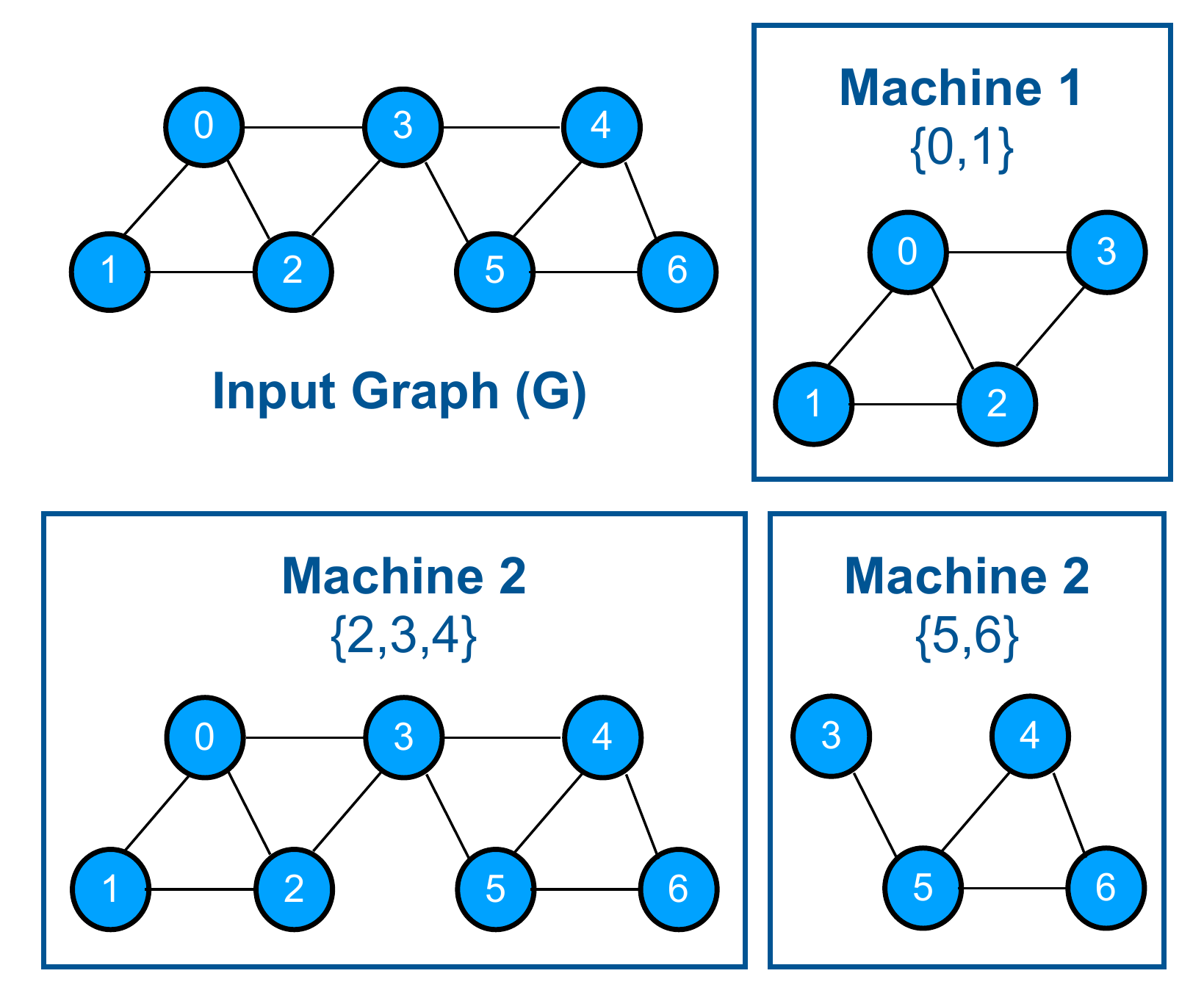}
    \caption{1-D Partitioned Graph}
    \label{fig:partition_gpm}
\end{figure}



\noindent
{\bf Moving computation to data}.
This approach, originated from Arabesque~\cite{teixeira2015arabesque}, 
performs the extension (computation) at the machine which 
keeps the required data. 
The partially-constructed embeddings during pattern enumeration 
are transferred among machines. 
For the example in Figure~\ref{fig:enum_tree} with 
graph partition of Figure~\ref{fig:partition_gpm},
starting from $v0$, machine 1
first locally extends it into three subgraphs
using local $N(0)$.
For the next extension, subgraph $(v0,v1)$ requires
$N(1)$, which is also local; while 
subgraph $(v0,v2)$ and $(v0,v3)$ require $N(2)$ and 
$N(3)$, which are stored in machine 2 according
to the partition. 
Thus, subgraph $(v0,v2)$ and $(v0,v3)$ are sent 
to machine 2, together with $N(0)$ since
it is needed for intersection between 
$N(2)$ and $N(3)$.
After machine 2 receives these data, it performs local extension. 
The extension of subgraph $(v0,v1)$ is performed 
in machine 1 locally. 

This approach has two drawbacks:
(1) excessive communication overhead due to 
the transfer of additional edge lists for computation; and 
(2) \red{difficult to reduce communication with data reuse.}
Its poor performance of 
\red{aDFS~\cite{trigonakis2021adfs}
on GPM confirms the drawbacks}. 


\noindent
{\bf Moving data to computation}.
This approach, which is used by G-thinker~\cite{yan2020g},
performs the embedding tree exploration from a given vertex $v$
in the machine that $v$ resides based on the input graph partition. 
For a graph with $|V|$ vertices, 
G-thinker performs enumeration with $|V|$ coarse-grained tasks
for the $|V|$ embedding tree explorations. 
For each tree, the task first fetches a k-hop subgraph \red{containing all data needed by its computation} and then 
performs computation only after the k-hop subgraph is fetched 
to local machine. 
\red{Thus, G-thinker can reduce communication by data reuse,
which is impossible for ``moving computation to data'' because the data transferred 
is partially-constructed subgraphs. }

The coarse-grained tasks of G-thinker limits the parallelism. 
Due to the high memory consumption of \red{the k-hop subgraph}, the number of tasks
(or trees) explored concurrently is small (about several hundreds).
Thus the capability that computation can be overlapped with communication
is limited, which leads to low CPU utilization. 
It is confirmed by our experiments:
\red{for triangle counting on the Patents graph, 
G-thinker's computation threads spend more than $70\%$ of runtime waiting for the communication thread to fetch remote data and update the software cache.}
More importantly, the map between task and data maintained by the 
cache incurs high overhead.

\begin{figure*}[t]
    \centering
    \includegraphics[width=\linewidth]{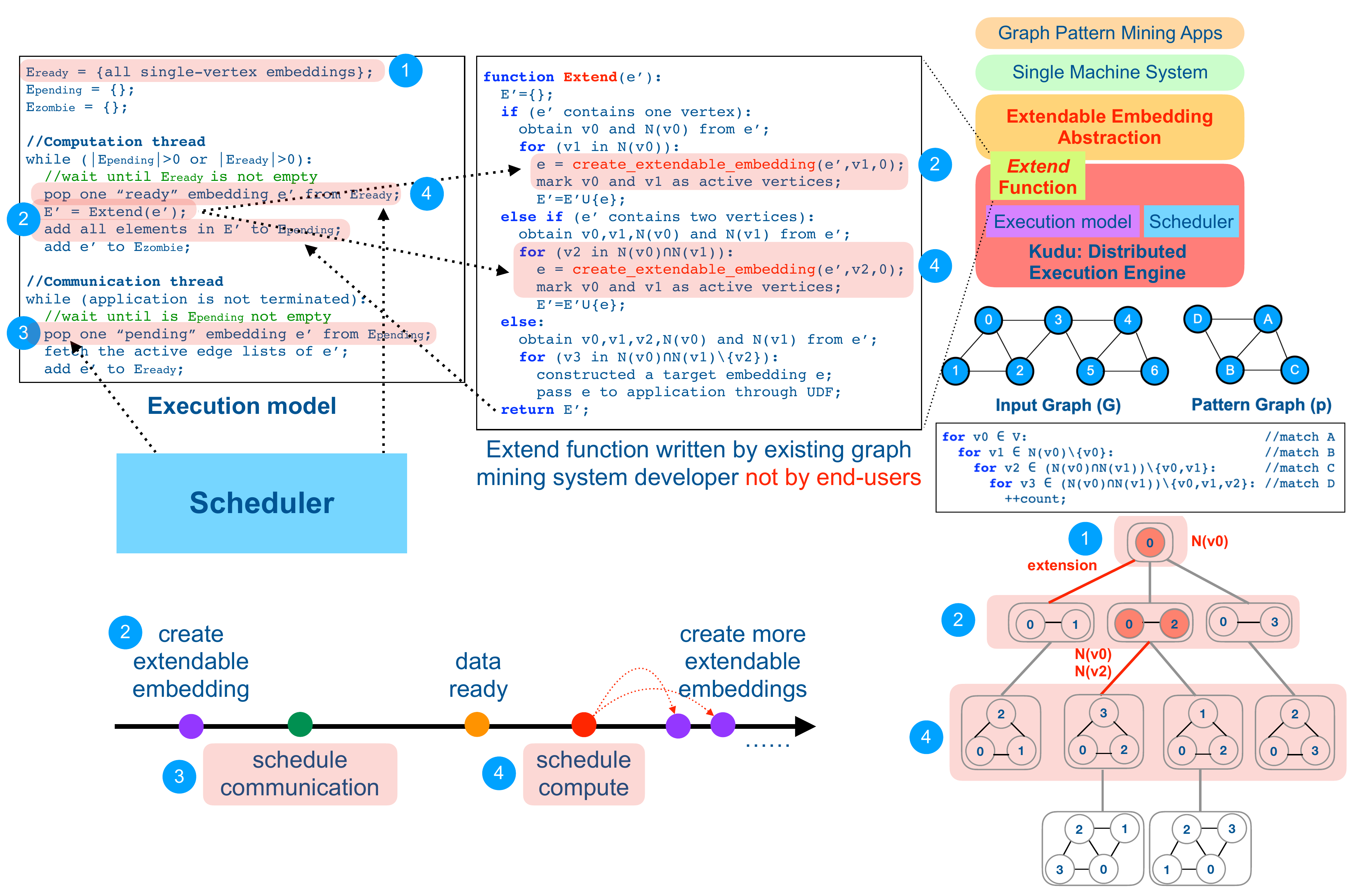}
    \caption{\red{Extendable embedding: concept, execution model and scheduler}}
    \label{fig:extendable_embedding_example}
\end{figure*}

%% file: extendable.tex
\section{Key Abstraction: Extendable Embedding}
\label{sec:extend}

\subsection{Definitions}
\label{sec:extend_def}

In pattern-aware enumeration, 
the embedding $e(p)$ matching a given pattern $p$ can be constructed by 
a sequence of extensions 
according to the pattern and the 
algorithm, which determines the order
of the extensions. 
To emphasize that these subgraphs
are enumerated in the process
of constructing $e(p)$, we
call the subgraphs enumerated in the 
sequence of extension also as embeddings (partially-constructed),
which match part of $p$.
Formally, each embedding $e(p)$ 
can be constructed by a 
sequence $e_0 \rightarrow e_1 \rightarrow ... \rightarrow e(p)$, where $e_0$ 
is a single vertex.

When $e_i$ is extended to $e_{i+1}$,
$e_i$ and $e_{i+1}$ are considered
to be the parent ($e_{parent}$) and 
child ($e_{child}$) embedding.
Based on the enumeration algorithm shown in Figure~\ref{fig:enum_tree},
extending $e_{parent}$ to $e_{child}$ 
requires the edge list of several 
{\em active vertices} in $e_{parent}$.  
The edge lists of active vertices 
are called active edge lists. 

{\em An extendable embedding is defined
as a partially-constructed embedding $e_i$ in a tree node,
along with active edge lists}.  
Without confusion, we can also
refer to an extendable embedding by $e_i$.
With the data of an extendable embedding available, $e_i$ can be 
extended to $e_{i+1}$.
The active vertices and edge lists
are determined by the pattern graph and 
the pattern enumeration algorithm. 
Based on our definition, 
the ``activeness'' follows the 
anti-monoticity property:
if a vertex in $e_i$ is inactive,
then it must be also inactive
in any $e_j$, where $j>i$.
This property enables 
the succinct storage of extendable
embeddings.
In Figure~\ref{fig:enum_tree}, the extendable embedding
for $(v0,v2)$ tree node is the subgraph itself with $v0$ and 
$v2$ being active vertices together with the edge lists
of the two vertices: $N(v0)$ and $N(v2)$.
Not all vertices in a partially-constructed embedding may be 
active. 
For example, the extendable embedding of tree node $(v0,v2,v3)$
has $v0$ and $v2$ (but not $v3$) as the active vertices and 
$N(v0)$ and $N(v2)$ as the active edge lists, because the next
extension only requires the intersection of $N(v0)$ and $N(v2)$. 

\subsection{System Interface}
\label{sys_interface}


Extendable embedding
is the key abstraction to
generate fine-grained tasks
that can be scheduled efficiently
leveraging the unique properties
of pattern-aware enumeration algorithms. Specifically, with graph partitioned
among distributed machines, 
when the data of an extendable embedding
$e_i$ are all locally available, 
the machine can perform the computation
to extend $e_i$ to $e_{i+1}$.
As a result, extendable embedding
breaks the embedding extension sequence
to generate each $e(p)$ into fine-grained
tasks with well-defined {\em dependent} data.


The {\em interface} exposed by \proj
to the client GPM systems is 
the \texttt{EXTEND} function. 
For a given pattern, the developer of a GPM system
needs to express the implementations with nested loops
in the \texttt{EXTEND} function based on the notion of 
extendable embedding. 
Figure~\ref{fig:extendable_embedding_example} shows the 
an example of the \texttt{EXTEND} function for the nested loops
enumerating pattern $p$ (on the right). 
The \texttt{EXTEND} function is called by 
the execution model (will be discussed next) with an extendable 
embedding as the parameter. 
Initially, all extendable embeddings are single vertices in the 
input graph. 

We can see that the correspondence between the nested loop
and the \texttt{EXTEND} function is intuitive:
the \texttt{EXTEND} function executes different branches
based on the current status of the extendable embedding.
Each branch: (1) performs the extension based on the locally 
available active edge lists (ensured by the execution model); 
(2) uses the \texttt{create\_extendable} \texttt{\_embedding()} function to create a new extendable embedding ;
(3) marks the active vertices and edges for the new extendable embedding;
and
(4) returns the new extendable embedding to the execution model,
which will be responsible to fetch the remote active edge lists
and schedule computation when they are ready.

The \texttt{create\_extendable\_embedding()} function is used to create an extendable embedding by adding one vertex to an existing embedding. The first two parameters of the function are the existing embedding and the new vertex. 
The third parameter represents the size of memory needed to store $e$'s reusable intermediate results (will be discussed in Section~\ref{sec:hierarchy}).
Essentially, the \texttt{EXTEND} function
breaks down the pattern enumeration
algorithm into fine-grained computation (intersection)
that can be flexibly scheduled by the execution model. 

For a compilation-based single-machine GPM system, the 
generation of the \texttt{EXTEND} function for different 
patterns can be done by modifying the compiler, which originally
generates the nested loop. 
We construct two distributed GPM systems by ``porting'' two compilation-based GPM systems, AutoMine~\cite{mawhirter2019automine} and GraphPi~\cite{shi2020graphpi},
to \proj in this manner. 
Thanks to the intuitive correspondence, the compiler modification
is very modest: {\em the porting effort is roughly 500 lines of codes
per system}.
With the \texttt{EXTEND} function specified, 
\proj engine can 
{\em transparently} orchestrate distributed 
execution.
The client GPM system developers just need
to focus on expressing pattern enumeration algorithm in 
the \texttt{EXTEND} function.


\subsection{Execution Model}
\label{sec:extend_exec}

\setlength{\intextsep}{2pt}%
\setlength{\columnsep}{8pt}%
\begin{figure}
    \centering
    \includegraphics[width=.9\linewidth]{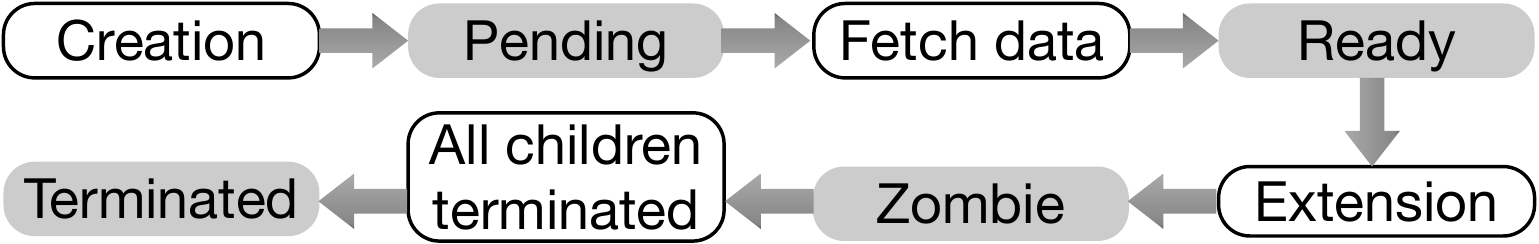}
    \caption{Extendable Embedding Execution States}
    \label{fig:extendable_embedding_life}
\end{figure}

The execution model maintains
four states for each extendable embedding, shown in Figure~\ref{fig:extendable_embedding_life}.
The ``pending'' state indicates that it has been created, but the active edge lists 
are not ready.
In this state, the computation is waiting for 
data. 
The ``ready'' state means that
the active edge lists have been fetched from either remote or local memory, and thus 
$e_i$ in the extendable embedding is 
ready to be extended to $e_{i+1}$.
The embedding extension is scheduled when
the computation resource is available. 
After the extension is performed, the state
changes to ``zombie'', which indicates that 
the extendable embedding's computation is 
finished, but its memory resource cannot
be released yet. 
It is because some data of the
completed extendable embedding
may be still shared with its children.
When all children of an extendable
embedding $e$ are completed,
the state is changed to ``terminated'', at which
point the system can release the memory
allocated to $e$.
Thus, the extendable embeddings are 
deallocated in a ``bottom-up'' fashion.

Figure~\ref{fig:extendable_embedding_example} (left)
shows the execution model of \proj.
It specifies the operations performed
by the computation and communication thread.
The details to ensure thread-safety
when multiple threads are used for 
computation and 
communication are discussed
in Section~\ref{sec:impl}.
The set
$E_{ready}$, $E_{pending}$, and 
$E_{zombie}$ contain 
the extendable embeddings in 
``ready'', ``pending'', and 
``zombie'' state, respectively.
At the beginning, $E_{ready}$ is initialized 
to contain all single-vertex embeddings, 
because the embedding enumeration of 
any pattern needs to start from them. 
The other two sets are initialized as empty sets. 
The computation thread keeps popping ``ready'' extendable embeddings from $E_{ready}$ and extending them to larger ones.
The \texttt{EXTEND} function will return a set of extended embedding 
$E'$ if it has not reached the
embedding of the pattern. 
Otherwise, the \texttt{EXTEND} function will 
return an empty set and call the user-defined
function (UDF) to pass the identified embedding to the
GPM application.
After returning from the \texttt{EXTEND} function,
newly generated extended embeddings in $E'$ 
are added to $E_{pending}$,
which will transparently 
trigger potential communication inside
the execution model.  
For the current embedding $e'$, since 
the extension has completed, it is 
added to $E_{zombie}$. 
We omit the details of 
keeping track of the children and 
state change from ``zombie'' to ``terminated''.

The communication thread runs continuously
until the end of application and 
executes whenever $E_{pending}$ is not 
empty. 
After popping one pending extendable 
embedding from $E_{pending}$,
it fetches the data, either locally or 
from a remote machine. 
For local data, the communication thread
just records the local data pointer.
The remote data fetch is blocking, but we
can batch multiple requests to 
amortize the network latency. 
After the communication is finished, 
the popped embedding from $E_{pending}$ 
is ready to be extended and added to 
$E_{ready}$,


\subsection{Running Example}
\label{run_example}

In this section, we describe several key steps in the \texttt{EXTEND}
function and execution model based on the pattern graph $p$ in 
Figure~\ref{fig:extendable_embedding_example}. 
Initially, $E_{ready}$ contains all single-vertex embeddings in
the local partition (\ding{182}) because the first extension 
step only needs the locally-available edge list these vertices.
The task of GPM is to explore the embedding tree for all vertices in 
the local graph partition.
The computation thread in the execution model will pick the 
extendable embeddings in $E_{ready}$ to perform extension.
In \ding{183}, the execution model takes an extendable embedding 
containing a single vertex $v0$ and its edge list $N(v0)$ and 
call the \texttt{EXTEND} function.
We also mark the steps in the embedding tree on the right.
Inside the \texttt{EXTEND} function, since $e'$ just contains
one vertex, the execution follows the first branch.
A new extendable embedding is created for 
the extendable embedding:
$v0$ extended with a neighbor vertex $v1$ corresponding to an edge 
in $N(v1)$.
Before returning from the \texttt{EXTEND} function,
$v0$ and $v1$ are marked as active vertices because 
$N(v0)$ and $N(v1)$ are needed for the next extension.
In this example, the returned $E'$ contains
three extendable embeddings:
$(v0,v1)$ with $N(v0)$ and $N(v1)$ as active edge lists;
$(v0,v2)$ with $N(v0)$ and $N(v2)$ as active edge lists; and
$(v0,v3)$ with $N(v0)$ and $N(v3)$ as active edge lists.
They are inserted to the $E_{pending}$.

Based on the graph partition in Figure~\ref{fig:partition_gpm},
$N(v2)$ and $N(v3)$ are stored in remote machine 2. 
Let us consider the extendable embedding 
$(v0,v2)$ with $N(v0)$ and $N(v2)$ as the example. 
Since $N(v2)$ is remote, the scheduler will later schedule
the communication thread at some point to fetch the data (popped from $E_{pending}$ in \ding{184}).
The communication is blocking, and when $N(v2)$ arrives at the 
local machine, the extendable embedding is inserted into 
$E_{ready}$, which is ready to be scheduled for computation (intersection)
by the computation thread.
For $(v0,v1)$ with $N(v0)$ and $N(v1)$, because both 
required edge lists are local, there will be no communication
incurred by the communication thread, and the computation 
will be scheduled directly.

In step \ding{185}, the computation thread is scheduled to 
process the extendable embedding 
$(v0,v2)$ with $N(v0)$ and $N(v2)$ in $E_{ready}$. 
Similar to \ding{183}, the execution model calls into 
the \texttt{EXTEND} function. 
This time, the extendable embedding contains two vertices 
$(v0,v2)$, so the execution falls into the second branch. 
New extendable embeddings are created for the extended 
embeddings found by the intersection of $N(v0)$ and $N(v2)$. 
In this example, two are generated:
$(v0,v2,v3)$ with $N(v0)$ and $N(v2)$ as active edge lists, and
$(v0,v2,v1)$ with $N(v0)$ and $N(v2)$ as active edge lists. 
Note that the active edge lists are the same since the next
extension still performs the intersection between them. 

On the second return from the \texttt{EXTEND} function, 
both $N(v0)$ and $N(v2)$ are local, so no communication is 
incurred to fetch $N(v2)$ again.
In \proj, it is realized by accessing the active edge lists
in parent node ($(v0,v2)$) through pointers in child nodes
($(v0,v2,v3)$ and $(v0,v2,v1)$).
The details are discussed in Section~\ref{sec:hierarchy}.

%% file: hybrid.tex
\section{Hybrid Embedding Exploration}
\label{sec:hybird}


\subsection{Motivation}
\label{sec:hy_motiv}



With DFS exploration, 
the extendable 
embeddings of the same tree
can be maintained
by a stack with small memory consumption.
The FILO order naturally satisfies the 
bottom-up memory release order. 
However, 
each tree only has one on-the-fly
extendable embedding at any time---significantly
limiting the capability of generating batched
communication. 
One may consider exploring multiple DFS trees concurrently to provide sufficient parallelism for communication batching and overlapping.
However, it does not solve the problem either.
Due to the power law, there are a small number of large DFS trees with high-degree starting vertices and many small trees. 
In the beginning, the system can generate batched requests and overlap communication and computation because of plenty parallelism.  Gradually, the DFSs of small trees quickly terminate, and there are only few large DFS trees left. Hence there are not sufficient batched requests and parallelism to hide the communication cost.

With BFS exploration, 
while sufficient number of 
on-the-fly extendable embeddings can be generated, the policy leads to inefficient 
memory management. 
The essential reason is that,
extendable embeddings are not released in the order that they are allocated. 
Thus, different objects 
vary in size and lead to 
the fragmentation problem.
In general, BFS tends to generate 
a very large number of embeddings---much larger than necessary for communication batching and overlapping---and results in enormous memory consumption.

\subsection{BFS-DFS Hybrid Exploration}
\label{sec:hy_disc}

A {\em chunk} is defined as a configurable number of 
extendable embeddings of the same level. 
We propose to
{\em apply DFS at a chunk granularity while exploring 
embeddings in a BFS manner within a chunk}. 
The memory of the chunk
can be allocated and deallocated together 
to avoid fragmentation. 
Moreover, we can control the memory
consumption with the chunk size. 
The data in the fixed memory for a chunk
can be horizontally shared 
among its extendable embeddings.
The lightweight mechanism to enable such
sharing is discussed in Section~\ref{sec:data_share}.

The hybrid BFS-DFS scheduling ensures that,
at any time, only one chunk of an embedding tree is being processed.
The memory consumption is bounded since 
we just need to allocate the {\em fixed memory for a chunk per level}
from the root to the current level being explored. 
On the other side, it eliminates the 
drawbacks of DFS since a chunk enables
many on-the-fly extendable embeddings within it,
which enables batched communication.



\begin{figure*}
    \centering
    \includegraphics[width=.92\linewidth]{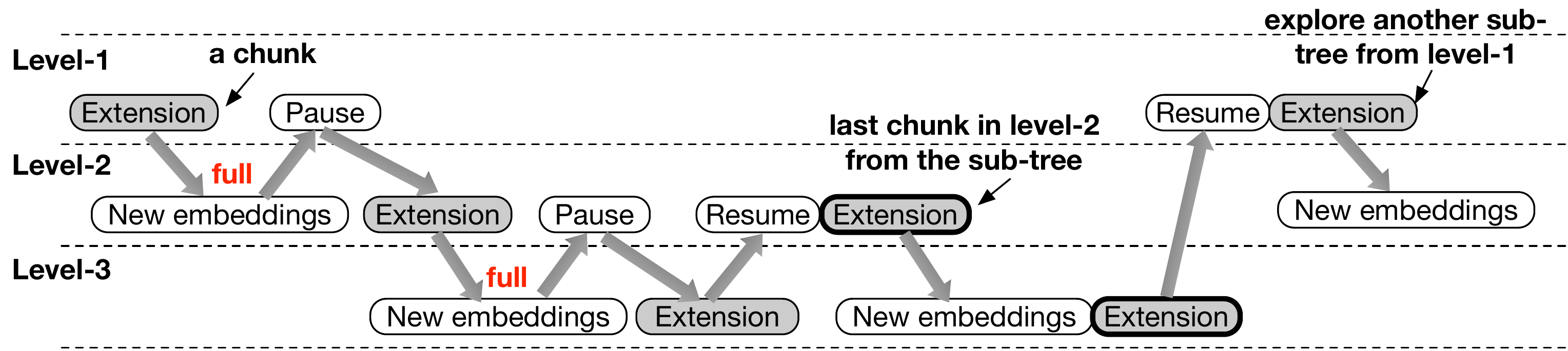}
    \caption{BFS-DFS Hybrid Exploration}
    \label{fig:hybrid_exploration}
\end{figure*}

During execution, the {\em current} chunk in tree level-$i$ performs
the extension and generates new extendable embeddings to be 
extended in the {\em next} chunk in tree level-$(i+1)$.
Before filling the next chunk, a fixed amount of memory (e.g., 1GB memory)
is pre-allocated. 
The next chunk is filled until the its memory is full. 
The procesure is shown in  Figure~\ref{fig:hybrid_exploration}.
At the beginning, the current level is $i$, so we only extend level-$i$'s embeddings until 
the memory for level-$(i+1)$ is full.
At this point, we stop the execution at level-$i$ and change level-$(i+1)$ as the current level.
Thus, the current level keeps going deeper in 
until it reaches the deepest level that does not generate any new embeddings, then it backtracks 
to the previous level. 
Once the current level changes from level-$(i+1)$
to level-$i$, all level-$(i+1)$'s extendable embeddings can be released because all their descendants have been processed, and the
execution of level-$i$ can resume.
The DFS at chunk granularity 
repeats until all embeddings are enumerated.

\subsection{Circulant Scheduling}
\label{sec:circ_schedule}

For a chunk, 
communication and computation
overlapping can be increased by circulant scheduling. 
As shown in Figure~\ref{fig:benefits} (right), on machine $K$, once a level becomes full, before
performing the extensions, the system shuffles all extendable embeddings at this level to $N$ batches according to the 
source machine ID in a circulant manner, where $N$ is the number of machines in the cluster.
These batches contain the extendable embeddings whose active edge lists reside on machine $K$, $(K+1)\%N$, $(K+2)\%N$, $\ldots$, $0$, $\ldots$, $(K+N-1)\%N$, respectively.
The key idea is to divide the execution of a
chunk into multiple steps, and
pipeline the execution and 
communication of the batches, so that 
in each step the computation of batch-$i$
is overlapped with the data fetch for batch-$(i+1)$.
A communication batch is the data transfer between 
the local machine and {\em one} particular remote machine. 

It is worth noting that we do not adopt 
the strict pipelining---the computation does not stall communication.
For example, once the data required by batch-$i$ has been fetched, the system immediately starts the communication of batch-$(i+1)$ 
without waiting for the completion of 
batch-$(i-1)$'s computation.



%% file: reduction.tex
\section{Low-Cost Data Sharing}
\label{sec:reduce}



\subsection{Vertical Data and Computation Sharing}
\label{sec:hierarchy}

\begin{figure}
    \centering
    \includegraphics[width=\linewidth]{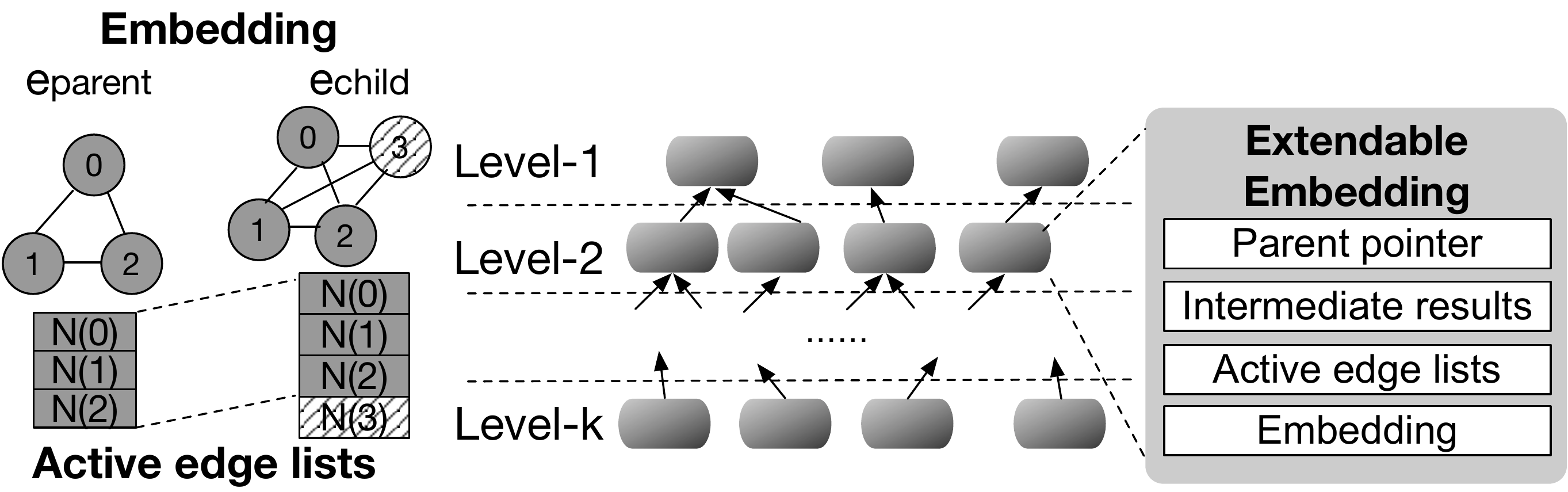}
    \caption{Vertical Data Reuse}
    \label{fig:hierarchical_representation}
\end{figure}

During embedding tree exploration,
different extendable embeddings may share common 
active edge lists.
Specifically, if an extendable embedding $e_{child}$ is extended from $e_{parent}$, 
most active edge list data 
of $e_{child}$ are also included in $e_{parent}$.
Figure~\ref{fig:hierarchical_representation} 
shows an example for 5-clique mining. 
Here, $e_{child}$ is obtained by extending $e_{parent}$ with $v3$,
which should be one of the vertices in the
intersection of $N(0)$, $N(1)$ and $N(2)$.
To extend $e_{child}$, $N(3)$ is needed.
Both $e_{child}$ and $e_{parent}$
include the active edge lists
$N(0)$, $N(1)$ and $N(2)$,
and $e_{child}$ only needs to 
additionally store $N(3)$ and refer $N(0) \sim N(2)$ to $e_{parent}$.
Since there is only one active vertex
of $e_{child}$ that is not included in $e_{parent}$, 
$e_{child}$ only needs to store the 
active edge list of one vertex.
Such data reuse reduces 
the communication as well: for an extendable 
embedding $e_{child}$, 
only the edge
list of the new vertex is fetched. 
Figure~\ref{fig:hierarchical_representation} (right)
shows the hierarchical representation of tree nodes in $k$ levels. 
The extendable embeddings in
level-$i$ have the same number of 
vertices, and point
to the embeddings in level-$(i-1)$.

\begin{figure}
    \centering
    \includegraphics[width=\linewidth]{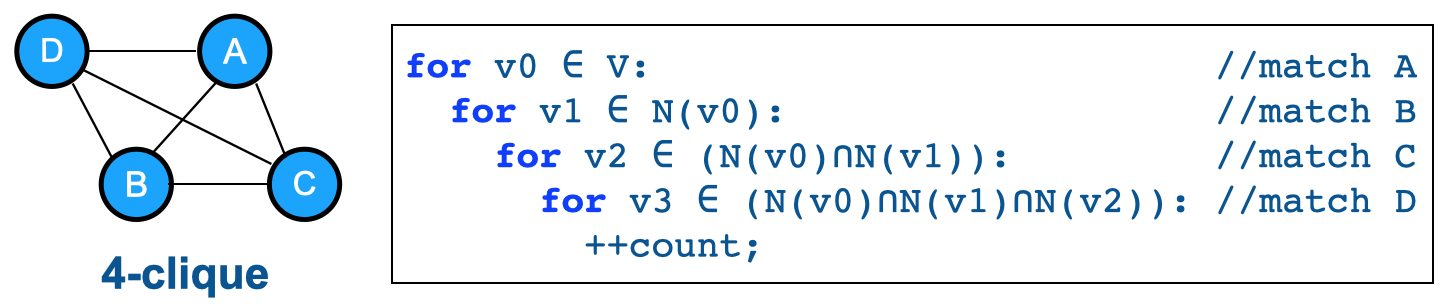}
    \caption{{Vertical Computation Reuse}}
    \label{alg:four_clique_mining}
\end{figure}

Extending an embedding $e$ may generate intermediate results that can be reused by its descendants to avoid computation redundancy~\cite{mawhirter2019automine}.
For 4-clique mining, shown in Figure~\ref{alg:four_clique_mining},
to extend an edge embedding $e_{edge}=(v_0,v_1)$ to a triangle $e_{triangle}$, we should find the common neighbors of $v_0$ and $v_1$, i.e., calculating $N(v_0)\cap N(v_1)$.
To further extend the triangle embedding $e_{triangle}=(v_0,v_1,v_2)$ to a 4-clique, 
we need to find those directly connected with $v_0$, $v_1$, and $v_2$, i.e., calculating $N(v_0)\cap N(v_1)\cap N(v_2)$.
In this case, extending $e_{triangle}$ can reuse the intermediate result $N(v_0)\cap N(v_1)$ of its parent $e_{edge}$ to avoid extra intersection cost.
The hierarchical data representation for vertical data reuse 
can naturally enable intermediate result sharing
between parent and child extendable embedding.
The intermediate results that can be shared
across levels are determined by algorithm
and can be indicated in the \texttt{EXTEND} function. 
These results are stored in an extendable
embedding, its children can directly
access them and copy them to its own 
embedding object. In this way, such 
intermediate results can be accessed by 
all descendants. 


\subsection{Horizontal Data Sharing}
\label{sec:data_share}


\red{The extendable embeddings in the same chunk 
share the edge lists. 
If any of them is fetched from remote machines, that
should be only done once. } 
We maintain the active edge list of $v$ requested
by extendable embeddings $e$ in a chunk in a
per-level hash table with $v$ as the key and
$e$ as the value.  
When an extendable embedding $e$ 
requesting a new active edge list of $v$
(not in $e$'s parent), we first check whether there is
already an entry for $v$ 
in the hash table.
If so (suppose the value is $e'$), 
we add an extra pointer in $e$ to $e'$, indicating that the new active edge list
requested by $e$ can be found in $e'$.
There is no need to fetch or 
allocate memory for it.
Otherwise, the pointer of $e$ is added
to the hash table using $v$ as the key.

To minimize the computation cost, we do not support collisions for hash table insertion---if the hash table entry $hash(v)$ is already occupied by another vertex $u$ with the same hash value, we simply drop the insertion of $v$ rather than building up a collision chain.
This simple policy 
leaves a small amount of redundant
communication (or additional memory)
that would have been 
eliminated (or saved) but significantly reduce 
hash table overhead.
We find that the simple design can still
drastically reduce the communication cost. 
For instance, it reduces the communication volume from 4.4TB to 33.8GB for 5-clique mining on the LiveJournal~\cite{leskovec2009community} graph.

\subsection{Static Data Cache}
\label{sec:cache}

Data accesses in GPM show long-term locality.
Due to the power law and skewness of graphs, 
the graph data (edge list) of some vertices are much more frequently accessed than others, and 
therefore contribute to a large portion of communication cost. 
For example, the most frequently accessed 5\% graph data for 3-motif mining on the UK~\cite{boldi2004webgraph} graph contribute to 93\% communication.
These ``hot-spot'' vertices with 
higher degree are included in more embeddings
during the enumeration.

We design an efficient {\em static} software graph data cache
shared by all chunks, across different chunks at different levels. 
The cache size is typically 5\%-15\% of the graph size per node. 
The cache is empty at the beginning. 
During embedding enumeration,
every time the system is about to fetch the graph data of a vertex $v$, 
it will query the cache first to see whether the data of $v$ has been cached. 
If so, it directly obtains the data from it.
Otherwise, if $v$'s degree is larger than a threshold (e.g., 64) and the cache is not full, the system will cache $v$'s data after fetching it through the network.
Once the cache is full, the system will no longer insert any data to it since 
we do not support cache eviction and replacement.

We make this design choice to 
make the cache as lightweight as possible. 
Our ``first accessed first cached with threshold'' policy
approximately caches the most frequent data---effectively capturing the skewed 
graph access characteristics. 
The no replacement policy works well for
the following reason. 
Assuming that graph data accesses 
are temporally uniformly distributed,
if a vertex $v$ is more frequently accessed than $u$ in the whole access history,
it is also more likely that the first access to $v$ is earlier than that of $u$.
Thus, the more frequently accessed data
have higher chance to be placed in the cache.

\subsection{NUMA-aware Support}
\label{sec:numa}

Modern clusters usually adopt the NUMA architecture~\cite{lameter2013overview}---the memory and processors are distributed to multiple sockets, and accessing remote-socket memory is more expensive.
To reduce and hide the cost of cross-socket accesses, 
the graph partition of a $M$-socket node is further divided into $M$ sub-partitions, 
and each of them is assigned to one socket.
We run the BFS-DFS hybrid exploration \textit{independently} on each socket based on the local sub-partition,
which enables circulant scheduling to hide the cost of 
accesses across both sub-partitions and partitions. 
Similarly,
static data cache is also uniformly divided into $M$ partitions and distributed to each socket.
\red{The cache partition on each socket is allowed to cache the graph data residing on a remote socket of the same node, in addition to remote node,  
and hence reduces cross-socket accesses. }

%% file: impl.tex
\section{Implementation}
\label{sec:impl}


\proj engine is implemented in C++ and has approximately 5000 lines of code. 
We implemented two scalable distributed GPM systems based on 
partitioned graph, k-Automine and k-GraphPi, by porting two single-machine
systems 
Automine~\cite{mawhirter2019automine}
and GraphPi~\cite{shi2020graphpi} \rev{(in its single-node mode)} on \proj, both of which are
state-of-the-art systems.
GraphPi supports distributed 
execution with replicated graph, 
which is compared with k-GraphPi.
The porting effort is roughly 500 lines of codes per system,
which is significantly less than building a new system from scratch.
Since Automine is not open-sourced, 
k-Automine is modified from our own Automine implementation AutomineIH (in house) that achieves comparable performance with the published results.

\noindent\textbf{Multi-threading support. } 
We leverage multiple computation threads to extend the embeddings in a chunk.
The workload is distributed dynamically. 
Once a batch of extendable embeddings become ready (Section~\ref{sec:circ_schedule}),
they are divided into multiple  mini-batches that are the basic workload distribution units (64 embeddings per mini-batch). 
Mini-batches are added to a lock-free workload queue and distributed to computation threads on demand. 
Embedding extension generates new extendable embeddings that are inserted to the next-level chunk.
We protect the insertion by a mutex.
To avoid lock contention,
each computation thread has a small local buffer (the size is a half of the L1-D cache) containing the generated embeddings.
Once the buffer is full, the embeddings are flushed to the next-level chunk.

\noindent\textbf{Communication subsystem. }
The communication subsystem is built on top of MPI. 
It consists of graph data requesting threads and responding threads (ratio $1:1$). 
The ratio between communication threads and computation threads is $1:3$.
Each communication thread runs on a dedicated CPU core to avoid thread scheduling overhead.


%% file: eval.tex
\section{Evaluation}
\label{sec:eval}

\subsection{Evaluation Methodology}

\noindent\textbf{Configuration. } 
Unless otherwise specified, our experiment environment is an 8-node cluster 
with a 
56Gbps InfiniBand 
network.
Each node has two 8-core Intel Xeon E5-2630 v3 CPUs (one CPU per socket) and 64GB DDR4 RAM.
The MPI library is OpenMPI 3.0.1. 

\begin{table}[htbp]
\centering
\scalebox{0.85}{
\begin{tabular}{c|c|c|c|c|c}
         \hline
         Graph & Abbr. & |V| & |E| & Max.Degree & Size\\
         \hline
         MiCo~\cite{elseidy2014grami} & mc & 96.6K & 1.1M & 1.4K & 9.1MB \\
         Patents~\cite{leskovec2005graphs} & pt & 3.8M & 16.5M & 0.8K & 154.9MB \\
         LiveJournal~\cite{leskovec2009community} & lj & 4.8M & 42.9M & 20.3K & 363.9MB \\
         \hline
         UK-2005~\cite{boldi2004webgraph} & uk & 39.5M & 0.94B & 1.8M & 7.3GB \\
         Twitter-2010~\cite{kwak2010twitter} & tw & 41.7M & 1.5B & 3.0M & 11.5GB \\ 
         Friendster~\cite{yang2015defining} & fr & 65.6M & 1.8B & 5.2K & 13.9GB \\
         \hline
         Clueweb12~\cite{callan2012lemur} & cl & 978.4M & 42.6B & 75.6M & 324.7GB \\ 
         UK-2014~\cite{boldi2004webgraph,boldi2011layered,boldi2004ubicrawler} & uk14 & 787.8M & 47.6B & 8.6M & 360.5GB \\
         WDC12~\cite{meusel2012web} & wdc & 3.5B & 128.7B & 95.0M & 984.9GB \\
         \hline
    \end{tabular}}
    \caption{Graph Datasets~\cite{snapnets}}
    \label{tab:datasets}
\end{table}

\noindent\textbf{Datasets. }
Table~\ref{tab:datasets} shows
the evaluated datasets, 
including three small-size graphs (mc-lj), 
three median-size datasets (uk-fr),
and three large networks (cl-wdc).
GPM applications take undirected graphs as input,
for directed datasets, the edge direction is simply ignored. 
All datasets are pre-processed to delete self-loops and duplicated edges.

\noindent\textbf{Evaluated applications. }
We use four categories of GPM applications. 
\textit{Triangle Counting (TC)} is a task that counts the number of triangle. 
\textit{$k$-Motif Counting ($k$-MC)} discovers the embeddings for all size-$k$ patterns. 
\textit{$k$-Clique Counting ($k$-CC)} counts the number of embeddings of the $k$-clique pattern (a complete graph with $k$ vertices).
\red{\textit{Frequent Subgraph Mining (FSM)} finds all labeled patterns whose supports (i.e., frequencies)~\cite{bringmann2008frequent} are no lower than a user-specified threshold.}


\subsection{Overall Performance}


\noindent\textbf{Comparing with distributed systems. }
We first compare \proj based systems (k-Automine and k-GraphPi) with G-thinker~\cite{yan2020g}, the {\em only}
state-of-the-art distributed GPM system with partitioned graph.
The results are presented in Table~\ref{tab:compare_with_dist}.
\red{We notice that G-thinker's performance is very low with two sockets per node due to its lack of NUMA support.
Hence, we also report the single-socket runtimes of G-thinker/k-Automine/k-GraphPi in parentheses and use them to calculate speedups}.
k-Automine and k-GraphPi on average outperform G-thinker by \avgspeedupkautominegthinker and \avgspeedupkgraphpigthinker (up to \maxspeedupkautominegthinker and \maxspeedupkgraphpigthinker), respectively. 
The speedup on Patents graph is very high because it is less-skewed.
In G-thinker, the cache management overhead cannot be effective amortized by graph data accessing time, leading to quite low performance.

Next, we compare our systems with GraphPi~\cite{shi2020graphpi}, the fastest distributed GPM systems based on replicated graph (Table~\ref{tab:compare_with_dist}). 
Surprisingly, except for 5-CC on MiCo, {\em k-GraphPi consistently delivers better performance than GraphPi even with the remote graph accessing overhead}.
The performance improvement is attributed to two reasons: 
1) \red{GraphPi's implementation adopts a complicated
task partitioning and distribution method that leads to large overhead}, 
and hence is slower on small-size workloads;
2) By decomposing the subgraph enumeration process into fine-grained embedding extension tasks, \proj can exploit strictly more parallelism than GraphPi that only parallelizes the first or first few loops of the subgraph enumeration process in a coarse-grained fashion. 
\red{For 3-MC, k-Automine is slower than k-GraphPi because of GraphPi's better pattern matching algorithm.}

\begin{table}[htbp]
    \centering
    \scalebox{0.74}{
    \begin{tabular}{c|c|c|c|c|c}
         \hline
         \multirow{2}{*}{App.} & \multirow{2}{*}{G.} & k-Automine & k-GraphPi & GraphPi & G-thinker \\
         & & (partitioned) & (partitioned) & (replicated) & (partitioned) \\ 
         \hline
         \#nodes & & 8 & 8 & 8 & 8 \\
         \hline
         \multirow{6}{*}{TC} & mc & 40.2ms (34.5ms) & 35.3ms (35.0ms) & 704.4ms & 2.2s (1.2s) \\
& pt & 221.2ms (361.1ms) & 225.0ms (363.9ms) & 6.7s & 124.7s (27.2s) \\
& lj & 706.8ms (1.2s) & 722.4ms (1.2s) & 9.8s & 904.9s (34.6s) \\
& uk & 705.5s & 706.2s & 1268.4s & CRASHED \\
& tw & 2293.1s & 2300.6s & 2886.5s & CRASHED \\
& fr & 84.1s & 78.5s & 169.2s & CRASHED \\
         \hline
         \multirow{6}{*}{3-MC} & mc & 57.6ms (67.8ms) & 56.4ms (44.6ms) & 1.5s & 2.2s (1.2s) \\
& pt & 363.1ms (642.1ms) & 289.2ms (459.1ms) & 13.8s & 34.6s (27.5s) \\
& lj & 1.6s (3.0s) & 847.3ms (1.4s) & 20.1s & 215.5s (36.7s) \\
& uk & 1.1h & 689.4s & 1,380.7s & CRASHED \\
& tw & 2.8h & 2309.9s & 3,032.1s & CRASHED \\
& fr & 194.0s & 82.4s & 388.5s & CRASHED \\
         \hline
         \multirow{6}{*}{4-CC} & mc & 293.8ms (485.0ms) & 299.9ms (478.8ms) & 844.0ms & 2.2s (2.3s) \\
& pt & 370.2ms (627.7ms) & 362.8ms (606.0ms) & 6.7s & 35.7s (26.5s) \\
& lj & 4.7s (8.6s) & 4.8s (8.7s) & 12.8s & 89.1s (37.7s) \\
& uk & 5.1h & 5.1h & 8.6h & CRASHED \\
& tw & 6.7h & 6.8h & TIMEOUT & CRASHED \\
& fr & 132.9s & 137.7s & 177.8s & CRASHED \\
         \hline
         \multirow{4}{*}{5-CC} & mc & 9.8s (16.0s) & 9.9s (16.1s) & 8.2s & 52.5s (52.1s) \\
& pt & 780.7ms (985.7ms) & 777.6ms (994.5ms) & 6.8s & 37.5s (27.6s) \\
& lj & 169.6s & 169.3s & 174.7s & CRASHED \\
& fr & 204.3s & 210.3s & 260.0s & CRASHED \\
         \hline
    \end{tabular}}
    \caption{Comparing with GraphPi/G-thinker (Timeout: 10h); Runtimes with one socket per node shown in parentheses}
    \label{tab:compare_with_dist}
\end{table}

\begin{table}[htbp]
    \centering
    \scalebox{0.8}{
    \begin{tabular}{c|c|c|c|c|c}
         \hline
         App. & Graph & k-Automine & AutomineIH & Peregrine & Pangolin \\
         \hline
         \#nodes & & 1 & 1 & 1 & 1 \\
         \hline
         \multirow{6}{*}{TC} & mc & 83.5ms & 52.3ms & 68.7ms & 56ms \\
& pt & 1.2s & 330.7ms & 1.1s & 289ms \\
& lj & 4.6s & 2.8s & 3.8s & 2.2s \\
& uk & 1.7h & 2.0h & 1.3h & 26.6s \\
& tw & 4.7h & 8.6h & 5.7h & 747.7s \\
& fr & 497.2s & 378.3s & 305.2s & 384.6s \\
         \hline
         \multirow{5}{*}{3-MC} & mc & 222.9ms & 160.3ms & 84.7ms & 288ms \\
& pt & 2.3s & 930.9ms & 1.7s & 1.5s \\
& lj & 12.0s & 8.9s & 4.6s & 29.2s \\
& uk & 9.1h & TIMEOUT & 1.3h & TIMEOUT \\
& fr & 1425.8s & 1206.8s & 316.1s & 1.8h \\
         \hline
         \multirow{4}{*}{4-CC} & mc & 1.7s & 1.2s & 1.8s & 2.8s \\
& pt & 2.0s & 381.0ms & 1.3s & 773ms \\
& lj & 32.3s & 31.3s & 49.6s & 54.7s \\
& fr & 852.6s & 570.3s & 1237.5s & OUTOFMEM \\
         \hline
         \multirow{4}{*}{5-CC} & mc & 66.1s & 46.8s & 78.0s & 132.0s \\
& pt & 3.4s & 408.4ms & 1.5s & 967ms \\
& lj & 1259.3s & 982.9s & 2076.6s & OUTOFMEM \\
& fr & 1435.1s & 900.2s & 3032.8s & OUTOFMEM \\
         \hline
    \end{tabular}}
    \caption{Comparing with Single-machine Systems}
    \label{tab:compare_with_single_machine_system}
\end{table}

\noindent\textbf{Comparing with single-machine systems. }
To show the efficiency of \proj, 
we further compare k-Automine's single-node performance with three state-of-the-art single-machine systems and report the results in Table~\ref{tab:compare_with_single_machine_system}.
We see that 
k-Automine achieves comparable performance with the baselines for most workloads.  
It is even faster than AutomineIH for TC/3-MC on the uk and tw graphs because of \proj's 
fine-grained parallelism. 
On the other side,
k-Automine is less effcient on pt.
\red{We analyze the inefficiency on pt in Section~\ref{sec:breakdown}.}
Pangolin is extremely fast for TC on the uk and tw graphs because of  orientation~\cite{chen2020pangolin}, a powerful \rev{algorithmic} optimization specifically targeting triangle and clique counting on skewed graphs, which is not adopted by Automine or Peregrine.
\rev{This optimization can be easily integrated into \proj as a pre-processing step. We do not enable the optimization here for fair comparisons with Automine and Peregrine.}

\begin{figure}[htbp]
    \centering
    \includegraphics[width=\linewidth]{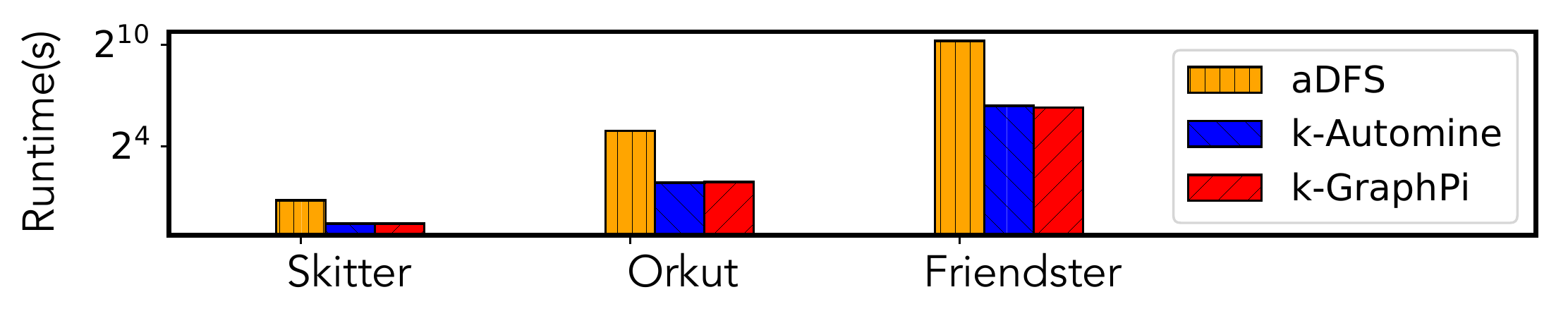}
    \caption{Comparing with aDFS}
    \label{fig:comp_adfs}
\end{figure}

\noindent\textbf{Comparing with graph-query systems.}
\red{We also compares our systems with aDFS in Figure~\ref{fig:comp_adfs}.
Since aDFS is not open-sourced, we compares to the TC runtimes reported in their paper~\cite{trigonakis2021adfs} on the Skitter~\cite{leskovec2005graphs}, Orkut~\cite{yang2015defining}, and Friendster~\cite{yang2015defining} graphs measured on a cluster with eight 28-core nodes.
Although with less computation resources (128 cores), 
k-Automine and k-GraphPi significantly outperforms aDFS by up to one order of magnitude thanks to our better ``move data to computation'' policy.}

\noindent\textbf{FSM performance.} 
{We ran k-Automine on both single-node and distributed modes on mc, pt, and lj graphs, and compared it with two single-node systems (Peregrine and AutomineIH) and a distributed system with 8-node (Fractal~\cite{dias2019fractal}).
We do not evaluate k-GraphPi or GraphPi since GraphPi does not come with a reference FSM implementation.
Similar to the Peregrine paper~\cite{jamshidi2020peregrine}, we only discover frequent patterns with no more than three edges.
Note that FSM requires labeled graphs. Hence, for unlabeled datasets like lj, we randomly synthesized their labels.
The results are shown in Table~\ref{tab:fsm_perf}.
k-Automine running in the single-node mode is consistently slower than AutomineIH.
The reason is that FSM processes a large number of candidate labeled patterns, 
and hence the per-pattern startup overhead of \proj (i.e., initializing the embedding execution engine) becomes non-negligible.
k-Automine running on 8 nodes is significantly faster than AutomineIH/Peregrine with one node and Fractal with 8 nodes thanks to its ability to leverage the \red{cluster-wide} computation resource efficiently.}

\begin{table}[htbp]
    \centering
    \scalebox{0.75}{
    \begin{tabular}{c|c|c|c|c|c|c}
         \hline
         Graph & Threshold & \multicolumn{2}{c|}{k-Automine} & AutomineIH & Peregrine & Fractal \\
         \hline
         \multicolumn{2}{c|}{\#Nodes} & 1 & 8 & 1 & 1 & 8 \\
         \hline
         \multirow{3}{*}{mc} & 3K & 91.1s & 18.8s & 46.7s & 181.3s & 105.1s \\
         & 4K & 90.1s & 15.6s
 & 45.3s & 170.6s & 102.1s \\
         & 5K & 89.7s & 15.2s & 45.1s & 166.9s & 108.3s \\
         \hline
         \multirow{3}{*}{pt} & 13K & 288.9s & 44.0s & 103.4s & 256.1s & 301.7s \\
         & 14K & 288.3s & 44.1s & 104.2s & 255.1s & 302.6s \\
         & 15K & 288.1s & 44.1s & 103.7s & 257.2s & 347.6s \\
         \hline
         \multirow{3}{*}{lj} & 800K & TIMEOUT & 1.9h & TIMEOUT & TIMEOUT & TIMEOUT \\
         & 900K & 6.3h & 0.8h & 5.7h & TIMEOUT & TIMEOUT \\
         & 1M & 6.3h & 0.8h & 5.7h & TIMEOUT & TIMEOUT \\
         \hline
    \end{tabular}}
    \caption{FSM Performance}
    \label{tab:fsm_perf}
\end{table}

\noindent\textbf{\rev{Performance on large-scale graphs.}}
We study \proj's scalability to large graphs by running k-Automine on three massive datasets (cl, uk14, wdc) for TC and 4-CC.
To the best of our knowledge,
wdc is the largest publicly available real-world graph with more than 3 billion vertices and 128 billion edges. 
k-Automine is evaluated on an 18-node cluster (two 16-core Intel Xeon-6130 CPUs and 128GB RAM each node).
Note that all of the three graphs cannot fit in the memory of a single node (128GB),
and hence cannot be processed by graph replication based systems like GraphPi.
For comparison purposes, we run AutomineIH for the same workloads on a high-end 64-core machine with two AMD Epyc-7513 CPUs and 1TB memory.
We adopt the triangle/clique-specific orientation optimization ~\cite{chen2020pangolin} mentioned previously, which converts the undirected input graph to a directed acyclic graph (DAG) as a pre-processing step to reduce the computation cost for both k-Automine and AutomineIH. 
The results are presented in Table~\ref{tab:perf_large_graphs}.
k-Automine significantly outperforms AutomineIH in all cases by up to $4.5\times$ ($3.2\times$ on average) thanks to \proj's ability to leverage a larger amount of cluster-wide computation resources.
With graph partitioning, \proj can efficiently scale to massive graphs
that requires the memory and computation resources 
\blue{beyond the capacity of a single machine}. 
More impressively, k-Automine even outperforms a state-of-the-art CPU-based distributed triangle counting implementation~\cite{pearce2017triangle} (with the orientation optimization) targeting large graphs, 
which counts all 9.65 trillion triangles of the wdc graph with 808.7 seconds using 256 24-core machines (in total 6144 cores). 
By contrast, k-Automine finished triangle counting on wdc with 200.5 seconds, achieving $4.0\times$ speedups with only $9.4\%$ of its computation resource (576 cores). 

\begin{table}[htbp]
    \centering
    \scalebox{0.95}{
    \begin{tabular}{c|c|c|c|c}
         \hline
         Graph & \#Vertices / \#Edges & App & k-Automine & AutomineIH \\
         \hline
         \multirow{2}{*}{cl} & \multirow{2}{*}{1.0B / 42.6B} & TC & 53.8s & 115.1s \\
         \cline{3-5}
         & & 4-CC & 0.75h & 2.6h \\ 
         \hline
         \multirow{2}{*}{uk14} & \multirow{2}{*}{0.8B / 47.6B} & TC & 94.1s & 299.4s \\
         \cline{3-5}
         & & 4-CC & 16.0h & 72.2h \\
         \hline
         \multirow{2}{*}{wdc} & \multirow{2}{*}{3.5B / 128.7B} & TC & 200.5s & 560.81s \\
         \cline{3-5}
         & & 4-CC & 18.8h & 66.5h\\
         \hline
    \end{tabular}}
    \caption{\rev{\proj's Performance on Large-Scale Graphs}}
    \label{tab:perf_large_graphs}
\end{table}

\subsection{Analyzing \proj Optimizations}


\noindent\textbf{Vertical computation sharing (VCS). } 
We run k-GraphPi for 4-CC and 5-CC with/without the optimization, 
and report  
\begin{figure}[htbp]
    \centering
    \includegraphics[width=\linewidth]{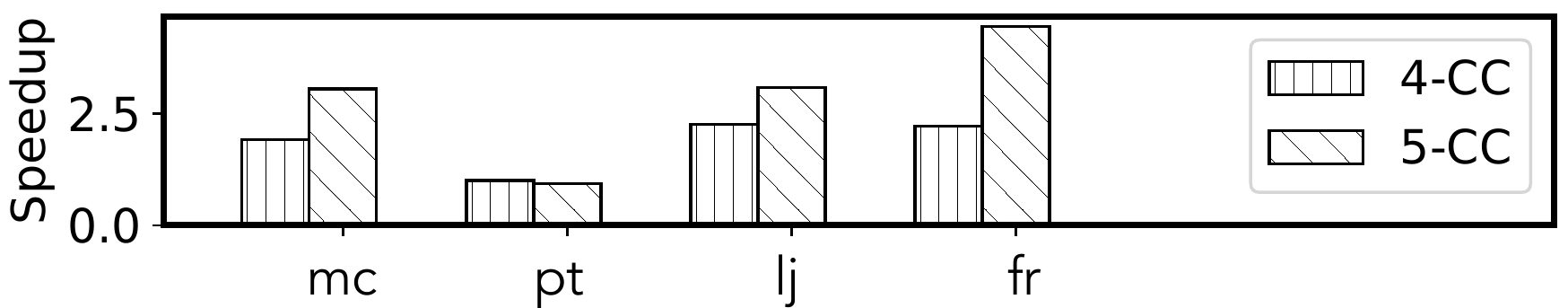}
    \caption{Speedup by VCS}
    \label{fig:speedup_vertical_comp_sharing}
\end{figure}
the
speedups in Figure~\ref{fig:speedup_vertical_comp_sharing}. 
The optimization improves
the performance 
by $2.10\times$ on average (up to $4.44\times$).
It is not very effective on the pt graph
since as mentioned earlier, embedding extension is lightweight on pt and only takes a small portion of execution time.

\noindent\textbf{Horizontal data sharing. }
We analyze the effect of horizontal data sharing (HDS) by running k-GraphPi for 4-CC and 
\begin{figure}[htbp]
    \centering
    \includegraphics[width=\linewidth]{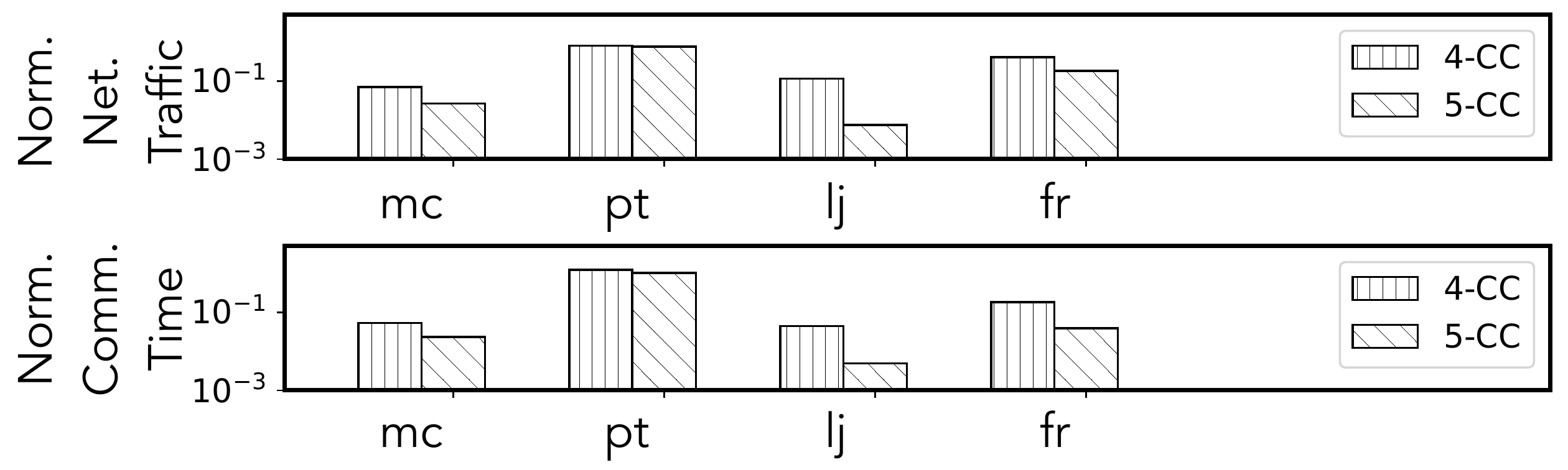}
    \caption{Effect of HDS}
    \label{fig:effect_communication_redundancy_elimination}
\end{figure}
5-CC with/without the optimization.
We report the network  
traffic
and communication time (normalized with respect to the version without the optimization) on critical path in Figure~\ref{fig:effect_communication_redundancy_elimination}. 
The optimization reduces network traffic and critical-path communication time by 70.5\% and 67.8\% on average (up to 99.3\% and 99.5\%), respectively.
The traffic reduction is moderate on the pt graph (20.4\% for 4-CC and 24.3\% for 5-CC).
It is because pt is less-skewed and thus
there are less ``hot-spot' active vertices appearing multiple times within a chunk.


\begin{table}[htbp]
    \centering
    \scalebox{0.95}{
    \begin{tabular}{c|c|c|c|c|c}
         \hline 
         \multirow{2}{*}{App.} & \multirow{2}{*}{G.} & \multicolumn{2}{c|}{Network Traffic} & \multicolumn{2}{c}{Runtime} \\
         \cline{3-6}
         & & with cache & no cache & with cache & no cache \\
         \hline
         \multirow{4}{*}{TC} & pt & 962.1MB & 1.0GB & 225.0ms & 228.5ms \\
& lj & 6.8GB & 7.9GB & 722.4ms & 770.8ms \\
& uk & 487.3GB & 57.7TB & 706.2s & 2615.2s \\
& fr & 1.4TB & 1.8TB & 78.5s & 89.6s \\
         \hline
         \multirow{3}{*}{4-CC} & pt & 1.2GB & 1.6GB & 362.8ms & 412.7ms \\
& lj & 15.7GB & 25.2GB & 4.8s & 4.9s \\
& fr & 2.3TB & 3.2TB & 137.7s & 185.3s \\
         \hline
         \multirow{3}{*}{5-CC} & pt & 1.3GB & 1.8GB & 777.6ms & 795.1ms \\
& lj & 33.8GB & 86.6GB & 169.3s & 169.5s \\
& fr & 2.7TB & 3.7TB & 210.3s & 250.1s \\
         \hline
    \end{tabular}}
    \caption{Analyzing the Static Data Cache (k-GraphPi)}
    \label{tab:static_data_cache}
\end{table}

\noindent\textbf{Static data cache. }
The effect of static data cache is reported in Table~\ref{tab:static_data_cache}.
The optimization significantly reduces network traffic
and hence improves end-to-end performance.
The optimization is extremely useful for highly-skewed graphs like uk. 
{\em For TC on uk, it reduces the traffic from 57.7TB to 487.3GB by more than 99\% (even with other optimizations like horizontal data sharing), and improves performance by $3.7\times$}.
Reduction in network traffic does 
not necessarily translate to performance benefit (e.g., 4-CC on lj) 
since communication cost is already completely hidden by computation.

\begin{table}[htbp]
    \centering
    \scalebox{0.95}{
    \begin{tabular}{c|c|c|c}
         \hline
         \multirow{1}{*}{App.} & \multirow{1}{*}{Graph} & With NUMA support  & No NUMA support \\ 
         \hline
         \multirow{3}{*}{4-CC} & pt & 2.1s (1.53x) & 3.2s\\
& lj & 33.0s (1.20x) & 39.5s\\
& fr & 870.4s (1.15x) & 998.5s\\
         \hline
         \multirow{3}{*}{5-CC} & pt & 4.0s (1.47x) & 5.9s\\
& lj & 1243.2s (1.02x) & 1269.5s\\
& fr & 1487.6s (1.30x) & 1930.7s\\
         \hline
    \end{tabular}}
    \caption{NUMA-aware Support}
    \label{tab:effect_numa_support}
\end{table}
\noindent\textbf{NUMA-aware support. }
We analyze our NUMA-aware support by running k-GraphPi on a single node. Table~\ref{tab:effect_numa_support} shows that
\proj's NUMA awareness leads to on average $1.26\times$ (up to $1.53\times$) performance gain.






\subsection{Scalability}

\begin{figure}[htbp]
    \centering
    \includegraphics[width=\linewidth]{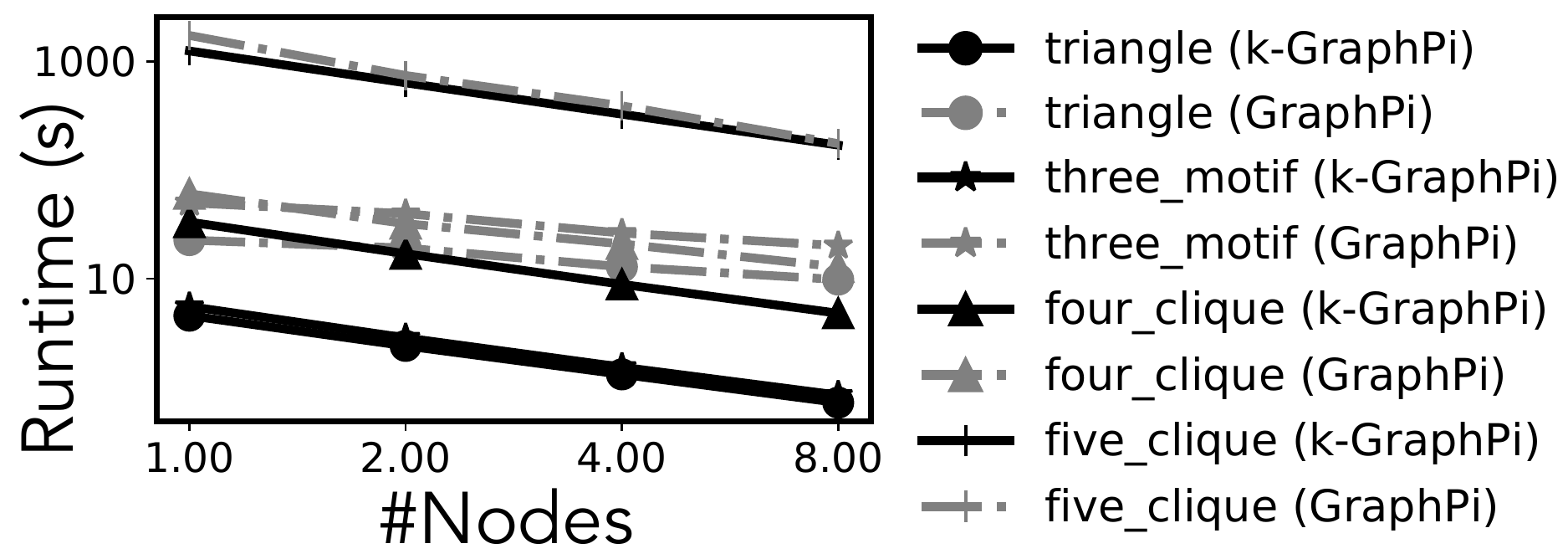}
    \caption{\proj's Inter-node Scalability (graph: lj)}
    \label{fig:inter_node_scalability}
\end{figure}

\noindent\textbf{Inter-node scalability. }
We report inter-node scalability of k-GraphPi and GraphPi in Figure~\ref{fig:inter_node_scalability} by varying the number of nodes.
k-GraphPi achieves similar or even better scalability compared with GraphPi,
and scales almost perfectly.
Leveraging 8 nodes is on average $6.77\times$ (up to $7.35\times$) faster than one node.
By contrast, GraphPi's speedup is on average $4.04\times$.

\begin{figure}[htbp]
    \centering
    \includegraphics[width=\linewidth]{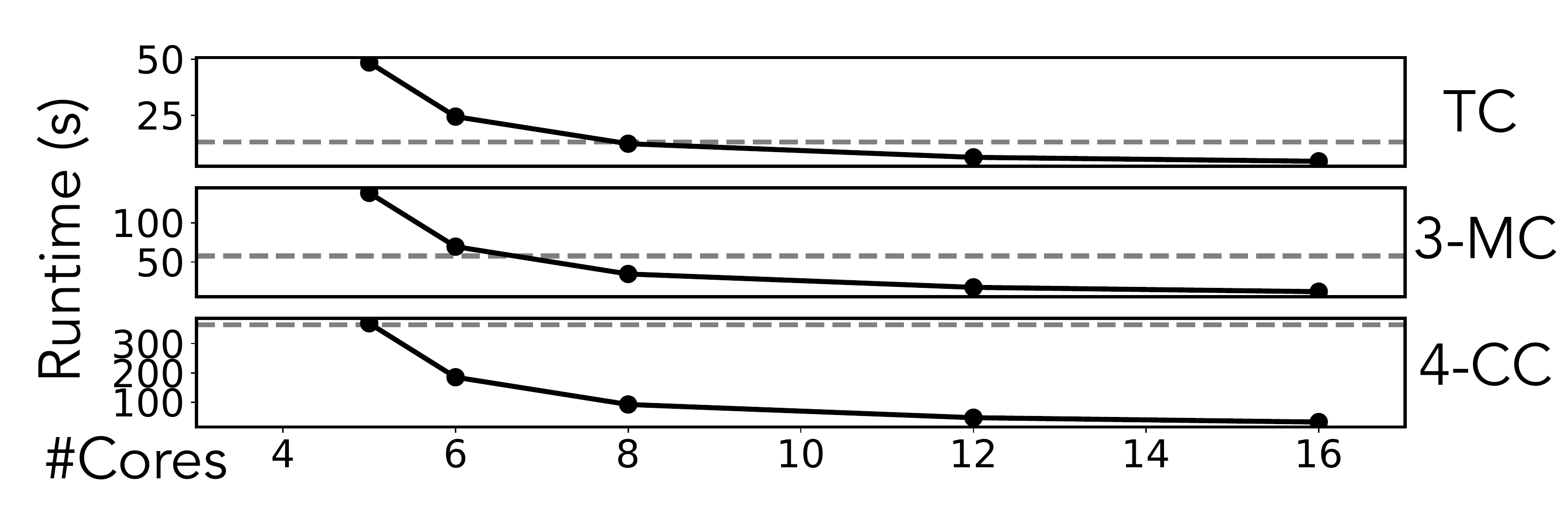}
    \caption{\revision{Intra-node Scalability}}
    \label{fig:intra_node_scalability}
\end{figure}

\noindent\textbf{Intra-node scalability and the COST metric. }
\revision{In Figure~\ref{fig:intra_node_scalability}, 
we analyze \proj's intra-node scalability and efficiency 
by running
k-Automine on a single node using
different numbers of CPU cores (5,}
\revision{
6, 8, 12, 16)
on graph lj.
The number of cores starts from five since four cores are always reserved for communication threads.
}
By utilizing \revision{16 cores}, k-Automine achieves $10.7\times$, $11.6\times$ and $11.4\times$ speedups for TC, 3-MC and 4-CC on the lj graph, respectively.
We also report the COST metric---the number of \revision{cores} with which a distributed system can outperform an efficient reference single-thread implementation~\cite{mcsherry2015scalability}.
We use the fastest single-thread runtime among Automine, Peregrine and Pangolin as the reference single-thread runtime (the dotted line in Figure~\ref{fig:intra_node_scalability}).
\revision{The COST metrics for TC, 3-MC and 4-CC are 8, 8, 6, respectively.}

\subsection{\revision{Execution Time Breakdown}}
\label{sec:breakdown}

\begin{figure}[htbp]
    \centering
    \includegraphics[width=\linewidth]{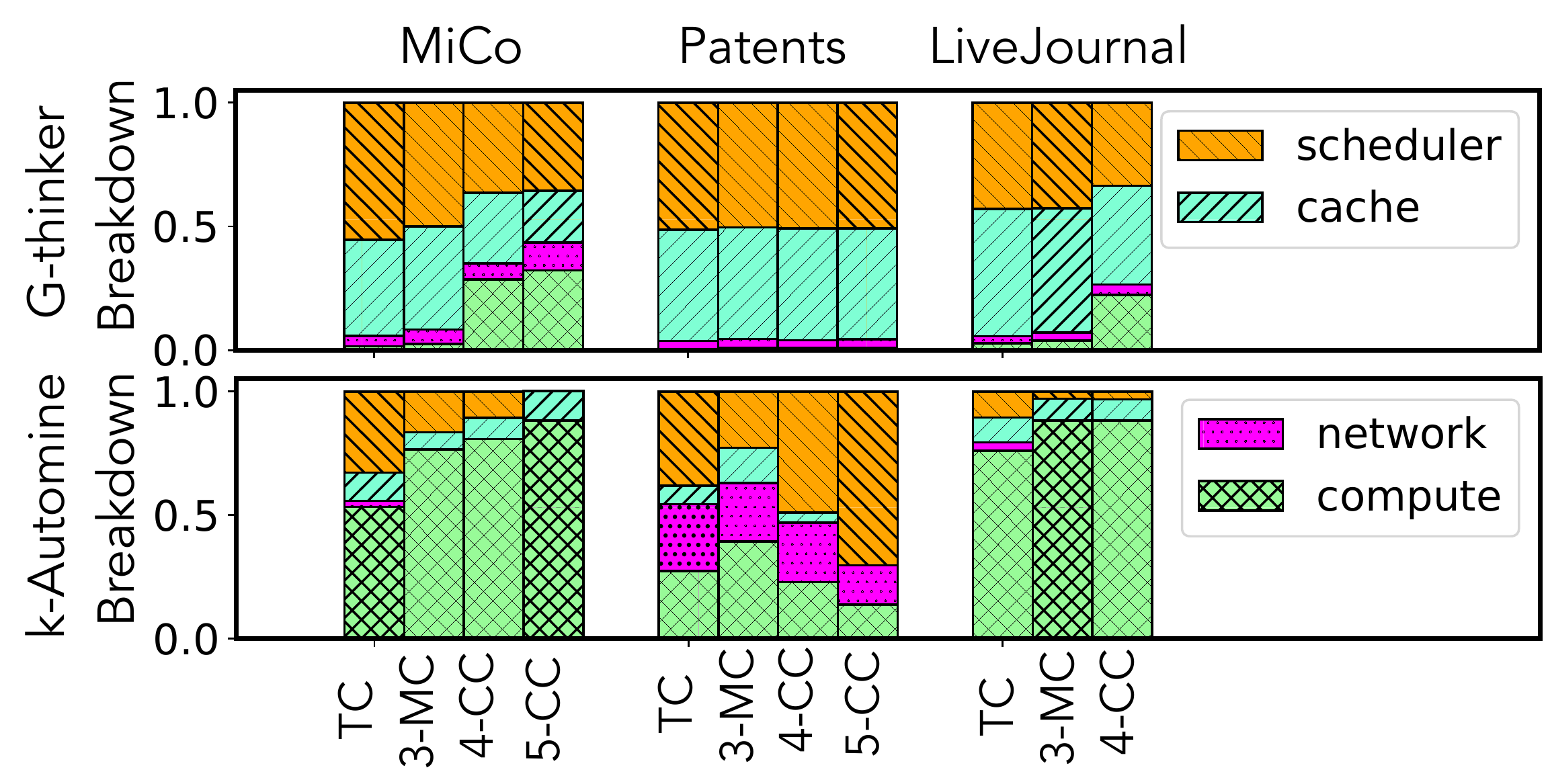}
    \caption{\revision{Runtime breakdown of G-thinker/k-Automine}}
    \label{fig:breakdown}
\end{figure}

\revision{
We compare the runtime breakdown of G-thinker and
k-Automine in Figure~\ref{fig:breakdown}.
There are missing data points (e.g., 5-CC on lj) since G-thinker crashes due to an internal bug. 
For G-thinker, most runtime is spent on cache maintaining (``cache'') (41.03\% on average) and scheduling (``scheduler'') (45.49\% on average). 
Only a small portion of time is spent on communication (``network'') (4.66\% on average) and embedding tree exploration (``compute'') (8.82\% on average).
By contrast, k-Automine significantly increases the portion of ``compute'' to 59.48\% on average (up to 88.24\%) thanks to our lightweight design. 
We also notice that except for Patents, 
the communication overhead of k-Automine is very low (less than 5\%), indicating the effectiveness of our communication optimizations.
The ``scheduler'' and ``network'' on Patents are still high (45.09\% and 22.70\% on average).
Patents is a less-skewed graph with low max-degree whose subgraph extension tasks are lightweight. 
Hence, the computation is not sufficient to amortize the scheduling overhead or hide the communication cost.}

\subsection{\revision{Cache Design Analysis}}

\begin{figure}[htbp]
    \centering
    \includegraphics[width=\linewidth]{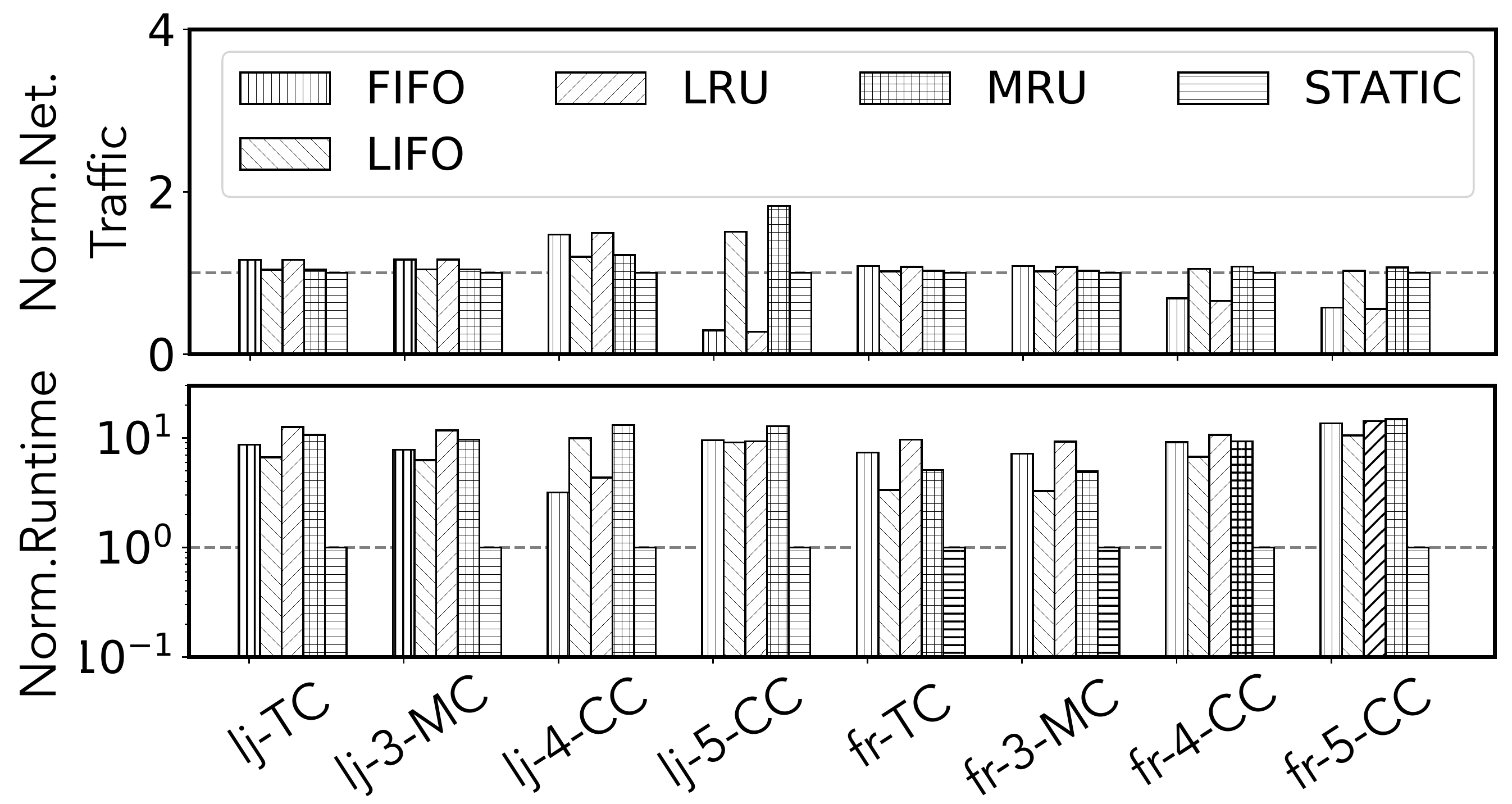}
    \caption{\revision{Comparing Different Cache Policies (k-GraphPi)}}
    \label{fig:various_cache_policies}
\end{figure}

\revision{
\noindent\textbf{Effects of cache policies.}
As shown in Figure~\ref{fig:various_cache_policies},
we evaluate different cache replacement policies, including FIFO (first-in-first-out), LIFO (last-in-first-out), LRU (least-recently-used), MRU (most-recently-used) and STATIC (no-replacement)
and report the volume of network traffic and execution time.
Both traffics and runtimes are normalized w.r.t. the STATIC policy.
In terms of traffic volume,
STATIC is similar or slightly better than LIFO and MRU
but worse than FIFO and LRU on three workloads (5-CC on lj and 4-CC/5-CC on fr).
It is because of STATIC's inability to capture the temporal change in data access pattern during execution---not a very surprising result because of its simple no-replacement design.
However, interestingly, more communication traffic on these workloads does not result in performance degeneration. 
We observed that in all cases, 
STATIC beats other policies by roughly one order of magnitude in terms of performance.
}

\revision{
The reason is twofold.
On the one hand, 
as shown in Figure~\ref{fig:breakdown},
communication is no longer a major bottleneck even with the suboptimal STATIC policy.
Further reducing the communication with a sophisticated policy is less beneficial as it has passed the point of diminishing returns.
On the other hand, 
policies with replacement incur more computation cost.
First, they have to maintain/update the cache during the whole execution while STATIC only needs to fill up the cache at the beginning of execution, completely eliminating the update cost after the cache is full.
Second, 
memory management is more complicated for policies with replacement. 
Note that the cached vertices have various degrees and need different sizes of memory blocks to store the graph data.
Hence, allocating/de-allocating the memory of a newly cached/uncached vertex cannot be done via a fast fix-size memory pool but requires a general-purpose dynamic memory allocator (e.g., UNIX malloc/free).  
Long-term frequent irregular memory allocation/de-allocation/re-allocation cause the fragmentation problem, 
significantly increasing the memory management cost.
In contrast, 
STATIC only needs allocations and thus avoids the fragmentation,
making its memory management much faster.
}

\begin{figure}[htbp]
    \centering
    \includegraphics[width=\linewidth]{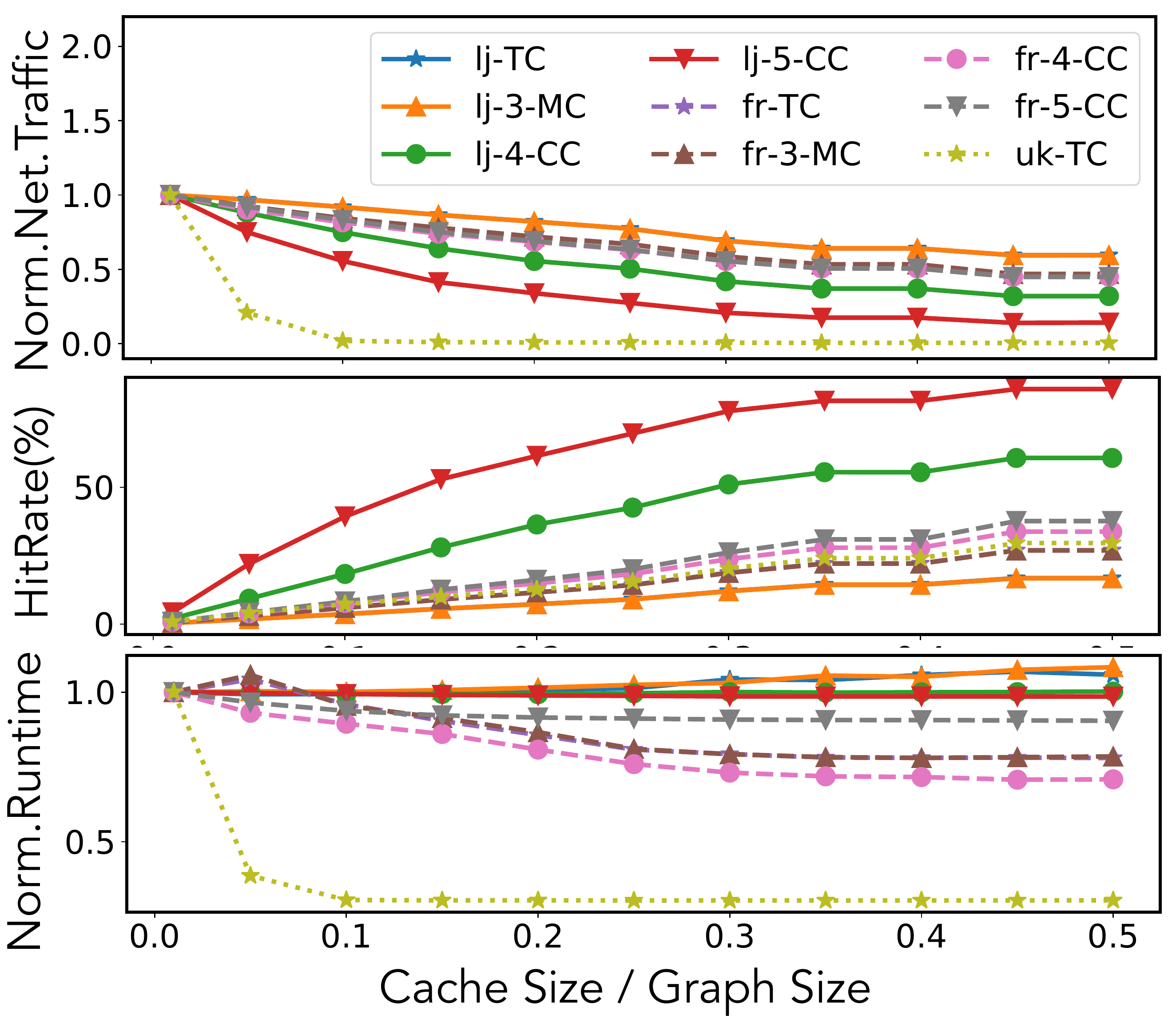}
    \caption{\revision{Varying Cache Size (k-GraphPi)}}
    \label{fig:various_cache_size}
\end{figure}

\noindent 
\revision{
\textbf{Effects of cache size.}
We evaluated k-GraphPi's performance with different cache sizes, 
varying from 1\%-50\% of the graph size, 
and reported the network traffic volumes, cache hit rates, and execution times in Figure~\ref{fig:various_cache_size}.
The traffic volumes and runtimes are normalized w.r.t. the cases when the cache size is 1\% of the graph size.
As the cache size increases, 
the runtime and traffic mostly decrease and the cache hit rate gets higher.
There is a point of diminishing returns where performance stops growing even with a larger cache (e.g., 10\% for uk-TC and 30\% for fr-4-CC)
as communication can be completely hidden with computation.
It is worth noting that the runtimes of lj-TC and lj-3-MC increase slightly (by 6-8\%) when the cache size increases to 50\%.
It is because of the effect of the hardware CPU cache. 
When the software cache size is small, it can be completely kept in the CPU L3 cache,
leading to low software-cache update/query overhead. 
As the size of the software cache grows, it exceeds the size of L3 cache, incurring higher update/query cost and hence affect the overall performance.}

\revision{
Although a larger cache can lead to higher performance, it affects the scalability to massive graphs:
if the single-node memory is 64GB and the cache size is set to 50\% of the graph size,
the system can only support graphs up to 128GB.
To achieve a good tradeoff between performance and scalabiltiy, 
we limit the cache size to be at most 15\% of the graph size for regular-size datasets and further reduce it to 3-4\% for massive datasets like WDC12.
}

\subsection{\revision{Sensitivity Tests on Chunk Size}}

\begin{figure}[htbp]
    \centering
    \includegraphics[width=\linewidth]{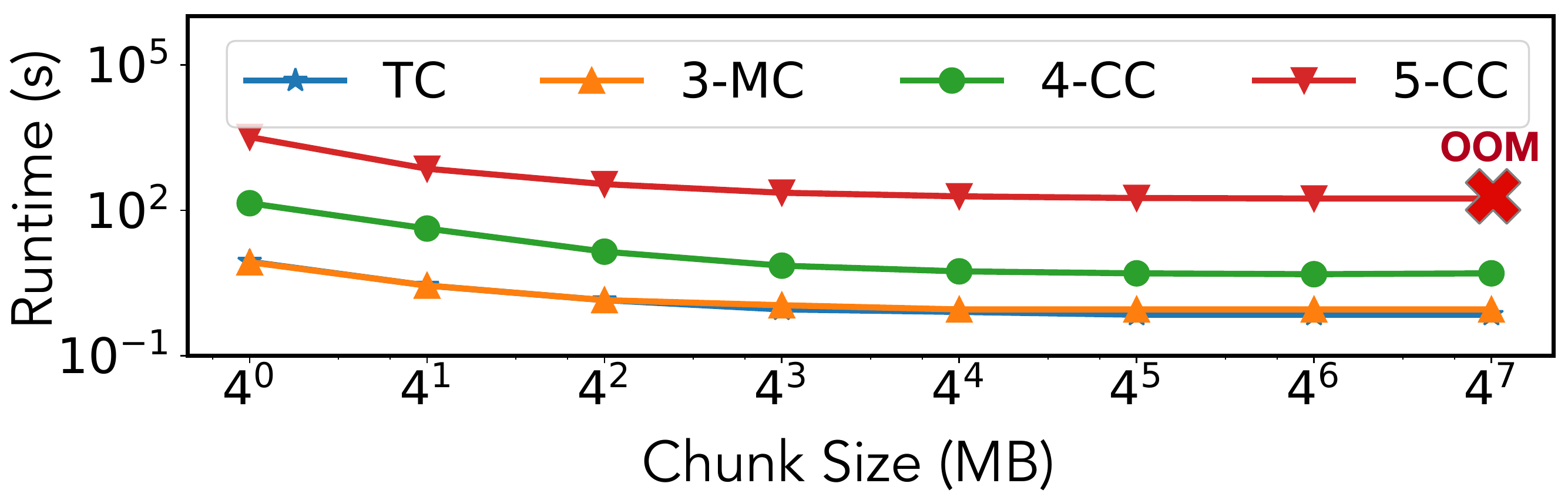}
    \caption{\revision{Varying Chunk Size (k-GraphPi, lj graph)}}
    \label{fig:various_chunk_size}
\end{figure}

\revision{We evaluate the performance with various chunk sizes (from 1MB to 16GB) in the hybrid exploration as shown in Figure~\ref{fig:various_chunk_size}.
The runtime decreases as the chunk size increases because of 1) more parallelism; 2) more opportunities to reuse data within a chunk. We empirically choose chunk size to be 4GB by default since it maximizes the performance while not introducing too much memory overhead (at most 4GB$\times$(5-1)=16GB for 5-CC). It is worth pointing out that the memory overhead is fixed regardless of graph size and will not affect the scalability. In general, the results suggest that the larger chunk is helpful, as long as the memory of the machine is sufficient.}

\subsection{\revision{Network Utilization}}

\begin{figure}[htbp]
    \centering
    \includegraphics[width=\linewidth]{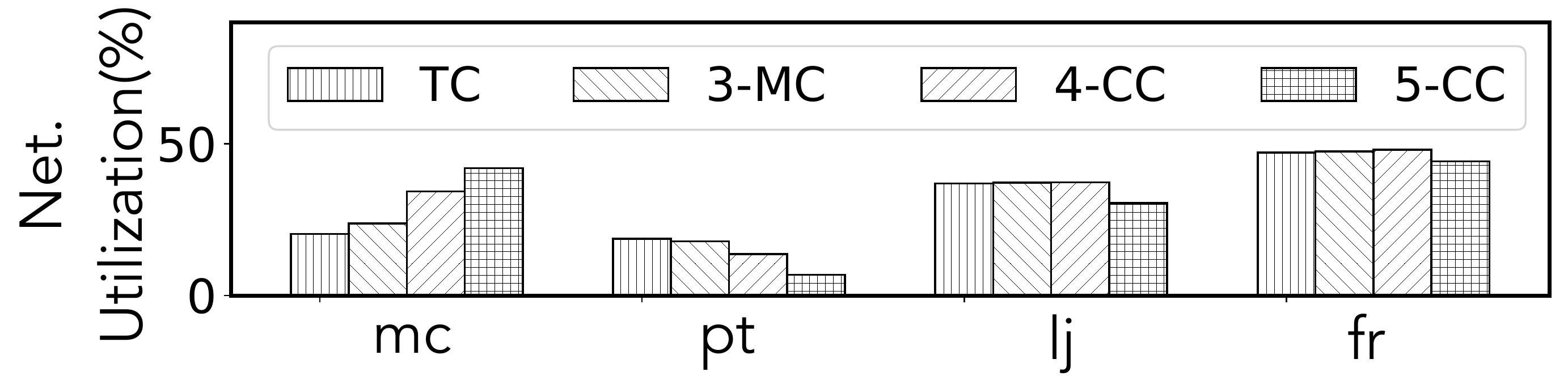}
    \caption{\revision{Network Bandwidth Utilization (k-GraphPi)}}
    \label{fig:net_utilization}
\end{figure}

\revision{
The network bandwidth utilization is shown in Figure~\ref{fig:net_utilization}. 
The system is compute-bounded on most workloads and thus network bandwidth is not saturated.
For most cases, the communication threads become idle to wait for more requests from computation threads. The only exception is the Patents graph, in which the communication thread serving data requests spent most of the time copying data from the local graph partition to a communication buffer so that they can be sent in a batch back to the requesting node. As a result, the network is largely underutilized. It is because most requests only access a small amount of data and have memory poor locality.
}

%% file: conc.tex
\section{Conclusion}
\label{sec:conc}

This paper proposes \proj, a distributed
execution engine 
that can be integrated
with existing single-machine graph pattern mining (GPM) systems.
The key novelty is the extendable embedding abstraction
which can express pattern 
enumeration algorithms, enable
fine-grained task scheduling,
enable low-cost
GPM-specific data reuse.
Two scalable distributed GPM systems 
are implemented by porting Automine and GraphPi on \proj.
Our evaluation shows that \proj based systems significantly outperform state-of-the-art distributed GPM systems with graph partitioning
\red{by up to \maxspeedupgthinker}, 
achieve similar or even better performance compared with the fastest graph replication based system,
and scale to \rev{massive graphs with more than one hundred billion edges}.